\documentclass[longbibliography,preprintnumbers,amsmath,amssymb,12pt]{revtex4-2}
\usepackage[caption=false]{subfig}
\usepackage{iftex}
\ifTUTeX
\usepackage{fontspec}
\else
\usepackage[T1]{fontenc}
\usepackage[utf8]{inputenc} 
\DeclareUnicodeCharacter{200B}{{\hskip 0pt}}
\fi
\usepackage[LGR,T1]{fontenc}
\usepackage{textgreek}
\usepackage{natbib}
\usepackage[english]{babel}
\usepackage[utf8]{inputenc}

\usepackage{enumitem}
\usepackage{braket}
\usepackage{footnote}
\usepackage{tablefootnote}
\usepackage[margin=1in]{geometry}
\usepackage{amsmath,amssymb,amsfonts,mathrsfs}
\usepackage{booktabs}
\usepackage{mathtools}
\usepackage{bm}
\usepackage{dcolumn}
\usepackage{graphicx}   
\usepackage{latexsym}
\usepackage{cmbright}
\usepackage[OT1]{fontenc}
\usepackage{kpfonts}
\usepackage[acronym]{glossaries}
\usepackage{physics}
\usepackage{siunitx}
\usepackage{comment}
\DeclareSIUnit{\persqrthz}{\ensuremath{\text{Hz}^{-1/2}}}
\usepackage[svgnames]{xcolor}
\usepackage[final]{hyperref}
\hypersetup{%
	colorlinks=true,
	citecolor=Blue,
}%

\newcommand{\myhyperref}[1]{\hyperref[#1]{\ref{#1}}}

\begin{document}

\title{Modelling the Sgr A$^*$ and M87$^*$ shadows by using the Kerr-Taub-NUT metrics in the presence of a scalar field}

\author{K. Jafarzade$^{1}$}
\email{khadije.jafarzade@gmail.com}

\author{M. Ghasemi-Nodehi$^{2}$}
\email{mghasemin@xao.ac.cn(Corresponding author)}

\author{F. Sadeghi$^{3}$}
\email{fatemeh.sadeghi96@ph.iut.ac.ir}

\author{B. Mirza$^{3}$}
\email{b.mirza@iut.ac.ir}

\affiliation{\footnotesize{$^{1}$ Department of Theoretical Physics, Faculty of Science, University of Mazandaran,
	P. O. Box 47416-95447, Babolsar, IRAN\\
	$^{2}$Xinjiang Astronomical Observatory, CAS, 150 Science1-Street, Urumqi 830011, China\\
	$^{3}$Department of Physics, Isfahan University of Technology, 84156-83111,
	Isfahan, Iran}}

\begin{abstract}
	\vspace{5mm}
	\begin{center}
		\textbf{Abstract}
	\end{center}
The recent unveiling of the images of Sgr A* and M87* has significantly advanced our understanding of gravitational physics. In this study, we derive a class of Kerr-Taub-NUT metrics in the presence of a scalar field (KTNS). Treating these metrics as models for supermassive objects, we constrain the parameters using shadow size estimates done by observations of M87* and Sgr A* from the Event Horizon Telescope (EHT). Comparing the obtained results with M87* data, we show an upper limit on the NUT charge $n$ such that the constraint on the shadow deviation from circularity ($ \Delta C $) will be fulfilled for $ n<0.5 $, and this allowed range changes with a variation in other parameters. Additionally, our findings reveal that fast-rotating KTNS metrics are better candidates for supermassive M87* than slowly rotating ones. We continue our study by estimating parameters using Keck and VLTI observations of Sgr A* and find that the constraint on the fraction deviation $ \delta $  is maintained within a certain range of the NUT charge such that the Keck bound is satisfied for $ n<0.41 $. In contrast,  the VLTI bound can be fulfilled for $ n>0.34 $. Finally, we investigate weak gravitational lensing using the Gauss-Bonnet theorem and illustrate that all model parameters increase the deflection angle,  causing light rays to deviate more significantly near fast-rotating KTNS objects.
\end{abstract}

    
\newacronym{e2e}{E2E}{End-To-End}
\newacronym{inrep}{INREP}{Initial Noise REduction Pipeline}
\newacronym{tdi}{TDI}{Time Delay Interferometry}
\newacronym{ttl}{TTL}{Tilt-To-Length couplings}
\newacronym{dfacs}{DFACS}{Drag-Free and Attitude Control System}
\newacronym{ldc}{LDC}{LISA Data Challenge}
\newacronym{lisa}{LISA}{the Laser Interferometer Space Antenna}
\newacronym{emri}{EMRI}{Extreme Mass Ratio Inspiral}
\newacronym{ifo}{IFO}{Interferometry System}
\newacronym{grs}{GRS}{Gravitational Reference Sensor}
\newacronym{tmdws}{TM-DWS}{Test-Mass Differential Wavefront Sensing}
\newacronym{ldws}{LDWS}{Long-arm Differential Wavefront Sensing}
\newacronym[	plural={MOSAs},
		        first={Moving Optical Sub-Assembly},
		        firstplural={Moving Optical Sub-Assemblies}
            ]{mosa}{MOSA}{Moving Optical Sub-Assembly}
\newacronym{siso}{SISO}{Single-Input Single-Output}
\newacronym{mimo}{MIMO}{Multiple-Input Multiple-Output}
\newacronym[plural=MBHB's, firstplural=Massive Black Holes Binaries (MBHB's)]{mbhb}{MBHB}{Massive Black Holes Binary}
\newacronym{cmb}{CMB}{Cosmic Microwave Background}
\newacronym{sgwb}{SGWB}{Stochastic Gravitational Waves Background}
\newacronym{pta}{PTA}{Pulsar Timing Arrays}
\newacronym{gw}{GW}{Gravitational Wave}
\newacronym{snr}{SNR}{Signal-to-Noise Ratio}
\newacronym{pbh}{PBH}{Primordial Black Holes}
\newacronym{psd}{PSD}{Power Spectral Density}
\newacronym{tcb}{TCB}{Barycentric Coordinate Time}
\newacronym{bcrs}{BCRS}{Barycentric Celestial Reference System}
\newacronym{lhs}{LHS}{Left-Hand Side}
\newacronym{rhs}{RHS}{Right-Hand Side}
\newacronym{mcmc}{MCMC}{Monte-Carlo Markov Chains}
\newacronym{cs}{CS}{Cosmic Strings}
\newacronym{ssb}{SSB}{Solar System Barycentric}
\newacronym{oms}{OMS}{Optical Metrology System}
\newacronym{dof}{DoF}{Degree of Freedom}
\newacronym{eob}{EOB}{Effective One-Body}
\newacronym{pn}{PN}{Post-Newtonian}
\newacronym{cce}{CCE}{Cauchy-Characteristic Evolution}
\newacronym{imr}{IMR}{Inspiral-Merger-Ringdown}
\newacronym{scird}{SciRD}{LISA Science Requirement Document}


%
\maketitle

\section{\label{Intro}Introduction}
One of the most interesting topics in general relativity and cosmology is the study of the final state of a star that has undergone continuous gravitational collapse
\cite{chiu1964gravitational, penrose1965gravitational, 2--penrose1969gravitational, hawking1970singularities, wald1971final, penrose1974gravitational, joshi2000gravitational}. 
The properties of the massive object that is created by continuous gravitational collapse depend on the properties of the star from which the collapse began. For example, depending on the initial data and the distribution of matter in the star, a singularity can be hidden at an event horizon or visible to external observers
\cite{christodoulou1994examples, harada2002physical, goswami2006quantum, joshi2009naked}. 
A final supermassive object can also be rotating or static, electrically charged or neutral, with or without a gravomagnetic monopole, with perfect geometric symmetry or with deviations from it, etc. If a homogeneous, static, neutral, spherically symmetric star collapses continuously, the final result is described by the Schwarzschild metric. In general, such conditions are very ideal and their occurrence is very rare, since most stars rotate around their axis, have inhomogeneous density, and do not have perfect geometric symmetry.

Naked singularities occur when the formation of the event horizon is delayed due to various factors, such as the non-uniform distribution of matter, which creates a region of very strong gravity that is visible to outside observers.
The cosmic censorship conjecture posits that naked singularities are not permissible
\cite{2--penrose1969gravitational, 3--hawking1980book};
however, this assertion remains unproven, and many singularities that occurs in general relativity, including the singularity of the Big Bang theory, are naked singularities
\cite{joshi2015story}. 
Since a region of strong visible gravity can provide an opportunity to observe the effects of quantum gravity, the study of naked singularities can be very useful. Also, the $\gamma$-metric (also known as Zipoy-Voorhees metric)
\cite{Darmois1927Les, 11--erez1959gravitational, 7--zipoy1966topology, 8--voorhees1970static}, 
a spacetime with naked singularities, has recently been proposed as a suitable candidate for describing Sgr A$^*$, and describes many of its features well, such as shadows and deviations from perfect sphericity
\cite{12--destounis2023geodesics, 13--lora2023q}.

A class of three-parameter metrics in the presence of a scalar field has recently been introduced
\cite{61--azizallahi2024three, 67--mirza2023class, 69--derekeh2024class, mirza2024exact, kachi2025class}, 
which contains the $\gamma$-metric and the Fisher-Janis-Newman-Winicour (FJNW) metric
\cite{fisher1999scalar, 9--janis1968reality, 10--wyman1981static}
for specific parameter values. In this paper, we first intend to derive the generalized form of these metrics in the presence of the rotation and NUT parameters. Given the rotation of most celestial bodies, finding the rotating form of the metrics is useful and closer to reality. Here, we will obtain the rotating and Taub-NUT form of these metrics for specific parameter values by using the Ernst equations 
\cite{ernst1968new, 64--ernst1968new}.
The NUT parameter also indicates the existence of a gravomagnetic monopole. A spacetime with a gravomagnetic monopole was first introduced in 1951 as a homogeneous cosmological model of the vacuum by Taub
\cite{taub1951empty}. 
This spacetime was also introduced again as a generalization of the Schwarzschild metric by Newman, Unti and Tamburino (NUT) in 1963
\cite{newman1963empty, kramer1980exact}. 
Metrics with a NUT parameter have some strange properties and their physical interpretation is not easy. So far, various interpretations of the Taub-NUT metric have been introduced. For some examples, see 
\cite{misner1963flatter, misner1967contribution, bonnor1969new, manko2005physical}. 
In 
\cite{69--derekeh2024class}, 
by using Ernst potential and Ehlers transformations  a class 
of metrics in the presence of a scalar field and a NUT parameter is introduced
\cite{ehlers1958konstruktionen, harrison1968new, ernst1976black}. 

The recent imaging of shadowy outlines of the supermassive compact entities Sgr A* and M87* by the Event Horizon Telescope (EHT) collaboration has initiated a novel avenue of inquiry within observational astronomy, aimed at exploring and evaluating the principles of gravity and fundamental physics in extreme gravitational fields
\cite{20--collaboration2019first, 21--akiyama2019event}. 
Consequently, we are approaching a point where we can investigate the characteristics of the geometry surrounding a astrophysical singularity and study how much these observations can distinguish well motivated models of Kerr-like metrics, emerging as solutions of physical theories, from general relativity. The characteristics of a shadow, including its form and size, are influenced by various metric parameters such as mass, angular momentum, and electric charge
\cite{22--de2000apparent}, 
as well as the properties of spacetime
\cite{23--johannsen2010testing, 24--cunha2017fundamental}
and the observer's location. The shadow image of the metrics functions as an invaluable resource for evaluating revised theories of gravity, while simultaneously offering critical insights into the behavior of jets and the dynamics of matter surrounding supermassive objects. The theoretical exploration of black hole shadows commenced with the work of Synge, who examined the shadow by Schwarzschild black holes
\cite{25--synge1966escape}. 
Following this, Bardeen and colleagues conducted an analysis of the shadow associated with Kerr black holes
\cite{26--bardeen1972rotating}. Their research indicates that the outline of the shadow formed by a non-rotating black hole is a flawless circle, whereas in the case of a rotating black hole, the shadow takes on an elongated appearance aligned with the rotation axis, a phenomenon attributed to the dragging effect. The phenomenon of shadows has been the subject of considerable research within the realm of black hole physics, such as regular black holes
\cite{27--jafarzade2022observational}, 
black holes within the framework of altered gravitational theory
\cite{28--amarilla2010null, 29--kumar2021shadows, 30--jafarzade2024thermodynamics}, black holes with scalar hair
\cite{31--khodadi2020black, 32--heydari2022null},  black holes within the context of non-linear electrodynamics
\cite{33--jafarzade2021shadow, 34--aliyan2024shadow}, 
black holes characterized by extra or higher dimensions
\cite{35--amir2018shadows, 36--nozari2024higher},  
the observations made by the EHT for applying rigorous limitations on the free parameters associated with black holes
\cite{37--davoudiasl2019ultralight, 38--hendi2023black, 39--jafarzade2024kerr, 40--ghasemi2020shadow, 41--kuang2022constraining, 42--ghasemi2024sgr, 43--ghasemi2021investigating}. The shadow of naked singularities are investigated and derived in
\cite{ortiz2015shadow, shaikh2019shadows, joshi2020shadow, dey2020perihelion, dey2021shadow, kaur2021comparing, shaikh2022shadows, solanki2022shadows, patel2022light, nguyen2023shadow, wang2023images, stuchlik2024shadows}.

The gravitational bending of light led to the first experimental confirmations of Einstein’s theory of relativity. Since that time, the subject of gravitational lensing has been explored not only in relation to singularities
\cite{44--iyer2007light, 45--virbhadra2009relativistic, 46--zschocke2011generalized} 
but also concerning various other entities, including wormholes and gravitational monopoles
\cite{47--abe2010gravitational, 49--deandrea2014editorial, 51--gibbons2008applications}.  
Currently, the phenomenon of gravitational lensing is extensively utilized as a significant method for investigating extrasolar planets, as well as dark matter and dark energy, within the realms of contemporary astronomy and cosmology
\cite{50--ishihara2016gravitational}. 
Gibbons and Werner devised a novel approach to examine the gravitational bending of light, leveraging the established Gauss-Bonnet theorem (GBT) as a basis. They then calculated the deflection angle of light in the framework of a spherically symmetric black hole spacetime
\cite{51--gibbons2008applications}. 
Following this, Werner employed the Finsler-Randers type optical geometry to apply this methodology to stationary black holes
\cite{52--werner2012gravitational}. 
Ishihara and colleagues assessed the angle of light deflection within static, spherically symmetric, and asymptotically flat spacetimes, considering the finite separation between a black hole, a light source, and an observer
\cite{53--ishihara2017finite}.
Later, Ono, Ishihara, and Asada 
\cite{54--ono2017gravitomagnetic} 
employed this technique for stationary and axisymmetric spacetimes. Subsequently this method was applied to study lensing in different singularity/wormhole geometries \cite{55--jusufi2018conical, 56--jusufi2018deflection, 57--crisnejo2018weak, 58--li2020circular, 59--li2020gravitational, 60--li2021kerr}.

This manuscript is structured as follows: In Sec. \ref{Sec1b}, we begin by presenting a specific class of three-parameter metrics that incorporate a scalar field. Subsequently, through the application of various transformations to the Ernst potential associated with these metrics, we derive the Kerr-Taub-NUT metrics within the context of a scalar field. In Sec. \ref{Sec2}, we analyze these solutions as supermassive entities and impose constraints on the metric parameters based on the bounds derived from the EHT observations of shadow characteristics. Sec. \ref{Sec4} is dedicated to the examination of the weak deflection angle at finite distances. The manuscript concludes with final observations in Sec. \ref{Sec5}.
\section{A class of Kerr-Taub-NUT metrics in the presence of a scalar field (KTNS)}\label{Sec1b}
In the following we review a class of solutions to Einstein's field equations that incorporate a scalar field. For our analysis, we posit that the scalar field exhibits spherical symmetry, while the metric maintains axial symmetry. The action under consideration is the Einstein-Hilbert action, which is minimally coupled to a massless scalar field, as follows
\begin{equation}\label{1}
\mathcal{S}=\int d^4 x ~\sqrt{-g}\left(R- \partial_{\sigma}\varphi(r)\partial^{\sigma}\varphi(r)\right).
\end{equation}
where $ R $, $ g $, and $ \varphi $ represent the Ricci scalar, the determinant of the metric, and the scalar field, respectively. The variation of the action (\ref{1}) concerning the metric tensor $ g_{\mu\nu} $ and the massless scalar field $ \varphi $ leads to
\begin{eqnarray}\label{metric}
&&R_{\alpha\rho}=\partial_{\alpha} \varphi(r)\, \partial_{\rho} \varphi(r),\\
&&\nabla_{\alpha} \,\nabla^{\alpha}\, \varphi(r)=0.
\end{eqnarray} 
A class of solutions to Einstein's equations characterized by the following three-parameter static metrics
\cite{61--azizallahi2024three}
\begin{equation}\label{2}
ds^2=-f^\gamma\, dt^2+f^\mu\, k^\nu\,\big(\frac{dr^2}{f}+r^2\,d\theta^{2}\big)+f^{1-\gamma}\,r^2\sin^2\theta\,d\phi^2,
\end{equation}
in which $\gamma$, $\mu$, and $\nu$ are parameters that satisfy the following constraint
\begin{equation}\label{mu}
\mu+\nu=1-\gamma
\end{equation}
Moreover, the functions $ f(r) $ and $ k(r,\theta) $ are defined as 
\begin{eqnarray}\label{15A}
&&f(r)  =1-\frac{2m}{r}, \\
&&k(r,\theta)=1-\frac{2m}{r}+\frac{m^2\sin^2\theta}{r^2}.
\end{eqnarray} 
The mass parameter, denoted as $ m $, is associated with the physical mass through the relationship $ M=\gamma m $. To obtain an expression for the scalar field $ \varphi $, we derive $ R_{rr} $ which leads to the following equation
\begin{equation}\label{7}
\varphi(r)=\sqrt{\frac{1-\gamma^2-\nu}{2}}~\ln\left(1-\frac{2m}{r}\right).
\end{equation}
From Eqs. (\ref{mu}) and (\ref{7}), we find that the following condition should be imposed on the parameters $\gamma$, $\mu$, and $\nu$ 
\begin{eqnarray}\label{9}
\mu\geq\gamma^2-\gamma,~~~ 1-\gamma^2\geq\nu.
\end{eqnarray}
Metric \eqref{2}, for the following parameters
\begin{eqnarray}\label{9a}
\mu=\gamma^2-\gamma,\qquad \nu=1-\gamma^2,
\end{eqnarray}
is known as the $\gamma$-metric and by selecting the parameters $\mu$ and $\nu$ as follows
\begin{eqnarray}\label{9b}
\mu=1-\gamma,\qquad \nu=0,
\end{eqnarray}
represents the FJNW metric. In the following we derive a class of Kerr-Taub-NUT metrics in the presence of a scalar field, for this purpose, we rewrite metric \eqref{2} for $ \gamma =1 $ in Lewis-Weyl-Papapetrou (LWP) metric form as below
\begin{equation}\label{h1}
ds^2=-f\,(dt-\omega\,d\phi)^2+f^{-1}\,\big[e^{2\,\lambda}\,(d\rho^2+dz^2)+\rho^2\,d\phi^2\big],
\end{equation}
that $ \rho $ and $ z $ are rewritten in terms of $ r $ and $ \theta $ as follows 
\begin{equation}\label{h2}
\rho=\sqrt{(m-r)^2-\sigma^2}\,\sin\theta, \qquad z=(r-m)\,\cos\theta,
\end{equation}
where $ \sigma=m $. Using Eq. \eqref{h2}, we rewrite metric \eqref{h1} in coordinates $ (t,\,r,\,\theta,\,\phi) $ and comparing the obtained result with metric \eqref{2} in $ \gamma=1 $, we get the following equations 
\begin{equation}\label{h3}
\begin{aligned}
&\omega=0,\\
&e^{2\,\lambda}=\big[1+\frac{m^2\,\sin^2\theta}{r\,(r-2\,m)}\big]^{\nu-1}.
\end{aligned}
\end{equation}
In 1968, Ernst wrote the Einstein-Maxwell equations as a summary with two equations
\cite{ernst1968new, 64--ernst1968new}, 
in the absence of an electromagnetic field, only one of these equations remains as follows
\begin{equation}\label{h5}
\varepsilon = f + i \, \chi.
\end{equation}
where $ \varepsilon $ is the Ernst potential and $ f $ is the coefficient $ dt^2 $ in metric \eqref{h1}. Also, the function $ \chi $ is defined as follows
\begin{equation}\label{h6}
\hat{\phi}\times\nabla\,\chi:=-\rho^{-1}\,f^2\,\nabla\,\omega,
\end{equation}
which $ \hat{\phi} $ is a unit vector in the direction of the $ \phi $ vector.
It has been shown in 
\cite{68--astorino2013embedding, astorino2015stationary, astorino2020enhanced}
that for the action in the presence of the scalar field, the Ernst equations do not change and only the equation related to the scalar field $ \Box\varphi=0 $ is added to the Ernst equations.

\begin{figure}[!htb]
\centering
\includegraphics[width=0.65\textwidth]{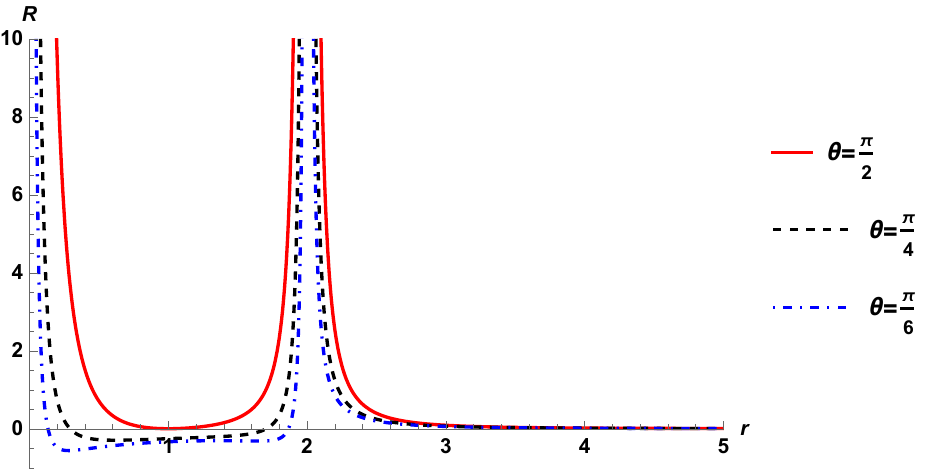}        
\newline
\caption{The Ricci scalar function for a class of KTNS metrics with parameters $ a=1 $, $n=1$, $m=1$, $\nu=-1$, diverge at $m+\sqrt{m^2-a^2+n^2} $ and $ m-\sqrt{m^2-a^2+n^2} $ for different values of $\theta$.}
\label{FigRicci}
\end{figure}

The Ernst potential corresponding to metric in Eq. \eqref{h1}  is equal to
\begin{equation}\label{h7}
\varepsilon=1-\frac{2\,m}{r}.
\end{equation}
By applying the following transformations to the Ernst potential \eqref{h7}
\begin{equation}\label{h8}
r\to r+i\,a\,\cos\theta+i\,n,\qquad m\to m+i\,n,
\end{equation}
we obtain a new Ernst potential in the presence of rotation ($ a $) and NUT ($ n $) parameters
\begin{equation}\label{h9}
\varepsilon^\prime=1-\frac{2\,(m+i\,n)}{r+i\,(a\,\cos\theta+n)}.
\end{equation}
For other methods of calculating Ernst potential in different cases, see
\cite{reina1976axisymmetric, 65--kinnersley1977symmetries, 66--kinnersley1977symmetriesII, kinnersley1978symmetriesIII, kinnersley1978symmetriesIV, hoenselaers1979symmetries, hoenselaers1979symmetriesVI, cosgrove1980relationships, wu2005two}. By considering Eq. \eqref{h9}, we obtain the functions $ f^\prime $ and $ \chi^\prime $ ($ \varepsilon^\prime=f^\prime+i\,\chi^\prime $) as follows
\begin{equation}\label{h10}
f^\prime=Re(\varepsilon^\prime)=1-\frac{2\,(m\,r+a\,n\,\cos\theta+n^2)}{r^2+a\,\cos\theta\,(a\,\cos\theta+2\,n)+n^2},
\end{equation}
\begin{equation}\label{h11}
\chi^\prime=Im(\varepsilon^\prime)=\frac{2\,\big[m\,a\,\cos\theta-n\,(r-m)\big]}{r^2+a\,\cos\theta\,(a\,\cos\theta+2\,n)+n^2}.
\end{equation}
Now, we derive the $ \omega $ function by using Eq. \eqref{h6}, where the gradient is written in the cylindrical coordinate system, and to convert the gradient to spherical coordinates, we should use the following equation 
\begin{equation}\label{h14}
\nabla\,h(r,\,\theta)=\frac{1}{\sqrt{(r-m)^2-m^2\,\cos^2\theta}}\;\big[\frac{\partial\,h(r,\,\theta)}{\partial\,r}\,\sqrt{r\,(r-2\,m)+a^2-n^2}\,\hat{r}+\frac{\partial\,h(r,\,\theta)}{\partial\,\theta}\,\hat{\theta}\,\big].
\end{equation}
Using Eqs. \eqref{h6} and \eqref{h14}, we have
\begin{equation}\label{h15}
\sqrt{r\,(r-2\,m)+a^2-n^2}\;\frac{\partial\,\chi^\prime}{\partial\,r}=\frac{f^{\prime\,2}}{\rho}\;\frac{\partial\,\omega^\prime}{\partial\,\theta},
\end{equation}
\begin{equation}\label{h16}
\frac{\partial\,\chi^\prime}{\partial\,\theta}=-\frac{f^{\prime\,2}}{\rho}\;\sqrt{r\,(r-2\,m)+a^2-n^2}\;\frac{\partial\,\omega^\prime}{\partial\,r}.
\end{equation}
In Eqs. \eqref{h15} and \eqref{h16}, $ \rho $ is determined using Eq. \eqref{h2} where the parameter $ \sigma $ is now defined by $ \sigma=\sqrt{m^2-a^2+n^2} $. Using Eqs. \eqref{h11} and \eqref{h16} we have
\begin{equation}\label{h18}
\omega^\prime=-\frac{2\,a\,(m\,r+a\,n\,\cos\theta+n^2)\,\sin^2\theta}{r^2-2\,m\,r+a^2\,\cos^2\theta-n^2}+C(\theta).
\end{equation}
By putting Eqs. \eqref{h18} and \eqref{h14} in Eq. \eqref{h15}, we obtain the function $ C(\theta) $ as follows
\begin{equation}\label{h19}
C(\theta)=-2\,n\,\cos\theta.
\end{equation}
Finally, according to Eqs. \eqref{h18} and \eqref{h19}, the function $ \omega^\prime $ is determined by the following equation
\begin{equation}\label{h20}
\omega^\prime=-2\,\frac{a\,(m\,r+n^2)\,\sin^2\theta+n\,(r^2-2\,m\,r+a^2-n^2)\,\cos\theta}{r^2-2\,m\,r+a^2\,\cos^2\theta-n^2}.
\end{equation}
Now we derive the $ \lambda $ and $ \varphi $ functions. For the simplicity of calculations in obtaining these functions, we go to $ (t,\,x,\,y,\,\phi) $ coordinates. The relations between coordinates $ r $ and $ \theta $ with $ x $ and $ y $ are as follows
\begin{equation}\label{h21}
r=\sigma\,x+m,\qquad\theta=\cos^{-1}y,
\end{equation}
where $ \sigma=\sqrt{m^2-a^2+n^2} $. Using Eqs. \eqref{h21}, we rewrite the Ernst potential in Eq. \eqref{h9} as follows
\begin{equation}\label{h22}
\varepsilon^\prime=1-\frac{2\,(m+i\,n)}{\sigma\,x+m+i\,a\,y+i\,n}.
\end{equation}
The derivative of the $ \lambda $ function in terms of $ x $ and $ y $ is determined by the following equations, respectively
\begin{equation}\label{h23}
\begin{aligned}
\partial_x\,\lambda^\prime=&\frac{1-y^2}{2\, \big[Re(\varepsilon^\prime)\big]^2\,(x^2-y^2)}\,\Big[x\,(x^2-1)\,\partial_x\,\varepsilon^\prime\,\partial_x\,\varepsilon^{\prime*}+x\,(y^2-1)\,\partial_y\,\varepsilon^\prime\,\partial_y\,\varepsilon^{\prime*}-y\,(x^2-1)\,(\partial_x\,\varepsilon^\prime\,\partial_y\,\varepsilon^{\prime*}\\
&+\partial_x\,\varepsilon^{\prime*}\,\partial_y\,\varepsilon^\prime)\Big]+\frac{1-y^2}{x^2-y^2}\,x\,(x^2-1)\,(\partial_x\,\varphi)^2,
\end{aligned}
\end{equation}
\begin{equation}\label{h24}
\begin{aligned}
\partial_y\,\lambda^\prime=&\frac{x^2-1}{2\,\big[Re(\epsilon^\prime)\big]^2\,(x^2-y^2)}\,\Big[y\,(x^2-1)\,\partial_x\,\varepsilon^\prime\,\partial_x\,\varepsilon^{\prime*}-y\,(1-y^2)\,\partial_y\,\varepsilon^\prime\,\partial_y\,\varepsilon^{\prime*}+x\,(1-y^2)\,(\partial_x\,\varepsilon^\prime\,\partial_y\,\varepsilon^{\prime*}\\
&+\partial_x\,\varepsilon^{\prime*}\,\partial_y\,\varepsilon^\prime)\Big]+\frac{(x^2-1)^2\,y}{x^2-y^2}\,(\partial_x\,\varphi)^2,
\end{aligned}
\end{equation}
where $ \varepsilon^{\prime*} $ is the complex conjugate of $ \varepsilon^\prime $. Considering the following equation
\begin{equation}\label{h25}
\partial_y\,\partial_x\,\lambda^\prime=\partial_x\,\partial_y\,\lambda^\prime,
\end{equation}
and using Eqs. \eqref{h23} and \eqref{h24}, we have
\begin{equation}\label{h26}
\partial_x^2\,\varphi^\prime + \frac{2 \, x}{x^2 - 1} \,\partial_x\, \varphi^\prime = 0, \quad \Rightarrow \quad \varphi^\prime = c_1 \, \ln \big( \frac{x - 1}{x + 1} \big) + c_2.
\end{equation}
According to the boundary conditions, the coefficients $ c_1 $ and $ c_2 $ are determined as follows
\begin{equation}\label{h27}
c_1=\sqrt{-\frac{\nu}{2}},\qquad c_2=0.
\end{equation}
By putting Eqs. \eqref{h22}, \eqref{h26} and \eqref{h27} in Eq. \eqref{h23} and then taking an integral of it with respect to $ x $ we have
\begin{equation}\label{h28}
\lambda^\prime=C^\prime+(\nu-1)\,\ln\big(x^2-y^2\big)-\nu\,\ln\big(x^2-1\big)+\ln\big[(m^2+n^2)\,(x^2-1)-a^2\,(x^2-y^2)\big].
\end{equation}
The Constant $ C^\prime $ can be derived by considering the boundary conditions. In the absence of the NUT parameter ($ n=0 $), the function $ \lambda^\prime $ corresponds to the Kerr metric in the presence of a scalar field
\cite{67--mirza2023class}, 
and in the absence of the rotation parameter ($ a=0 $), the function $ \lambda^\prime $ corresponds to the Taub-NUT metric in the presence of the scalar field
\cite{69--derekeh2024class}, 
so, constant $ C^\prime $ can be derived as follows
\begin{equation}\label{h29}
C^\prime=-\ln\big(m^2-a^2+n^2\big).
\end{equation}
Using the following coordinate transformation
\begin{equation}\label{h30}
x=\frac{r-m}{\sigma},\qquad y=\cos\theta.
\end{equation}
And according to the Eqs. \eqref{h26}, \eqref{h27}, \eqref{h28} and \eqref{h29}, the functions $ e^{2\lambda^\prime} $ and scalar field $ \varphi^\prime(r) $ can be written as follows 
\begin{equation}\label{h31}
e^{2\lambda^\prime}=\frac{r^2-2\,m\,r+a^2\,\cos^2\theta-n^2}{(m-r)^2-(m^2-a^2+n^2)\,\cos^2\theta}\;\big[1+\frac{(m^2-a^2+n^2)\,\sin^2\theta}{r^2-2\,m\,r+a^2-n^2}\big]^\nu,
\end{equation}
\begin{equation}\label{h32}
\varphi^\prime(r)=\sqrt{-\frac{\nu}{2}}\,\ln\left(\frac{r-m-\sqrt{m^{2}-a^{2}+n^2}}{r-m+\sqrt{m^{2}-a^{2}+n^2}}\right).
\end{equation}
By inserting Eqs. \eqref{h10}, \eqref{h20} and \eqref{h31}, into Eq. \eqref{h1}, we obtain a class of Kerr-Taub-NUT metrics in the presence of a scalar field as follows
\begin{equation}\label{metric19}
\begin{aligned}
ds^2&=-\frac{\Delta-a^{2}\,\sin^{2}\theta}{\mathcal{R}^{2}}\,dt^{2}+\frac{\mathcal{R}^{2}}{\Delta}\,\left(1+\frac{(m^{2}-a^{2}+n^{2})\,\sin^{2}\theta}{\Delta}\right)^{\nu}\,dr^{2}\\
&+\mathcal{R}^{2}\,\left(1+\frac{(m^{2}-a^{2}+n^{2})\,\sin^{2}\theta}{\Delta}\right)^{\nu}\,d\theta^{2}+\frac{\left(r^{2}+a^{2}+n^{2}\right)^{2}\,\sin^{2}\theta-\Delta\,\big(a\,\sin^{2}\theta -2\, n\, \cos \theta\big)^{2}}{\mathcal{R}^{2}}d\,\phi^{2}\\
&+\frac{2\,\big(a\,\sin^{2}\theta -2\, n\, \cos \theta\big)\,\Delta-2\,a\,\sin^{2}\theta\,(r^{2}+a^{2}+n^{2})}{\mathcal{R}^{2}}\,dt\,d\phi,
\end{aligned}
\end{equation}
where
\begin{eqnarray}
\Delta &=& r^{2}-2\, m\, r +a^{2}-n^{2}, \nonumber\\
\mathcal{R} &=& \sqrt{r^{2}+(a\,\cos \theta+n)^{2}}, \nonumber
\end{eqnarray}
In the limit of $ a=0 $, metric (\ref{metric19}) reduces to TNS metric obtained in
\cite{69--derekeh2024class}. 
Now, we are going to investigate the singularities of a class of KTNS metrics. For this purpose, we use Ricci scalar. Ricci scalar related to metric \eqref{metric19} is calculated as follows
\begin{equation}\label{RK1}
R=-\frac{2\,\nu\,(m^2-a^2+n^2)}{(r^2-2\,m\,r+a^2-n^2)\,\big(r^2+a\,\cos\theta\,(a\,\cos\theta+2\,n)+n^2\big)}\;\Big[\frac{(m-r)^2-(m^2-a^2+n^2)\,\cos^2\theta}{r^2-2\,m\,r+a^2-n^2}\Big]^{-\nu},
\end{equation}
Due to the negativity of parameter $\nu$, the Ricci scalar function has singularities in the following points
\begin{equation}\label{RK1a}
r_{\text{singularity}}=m\pm\sqrt{m^2-a^2+n^2}.
\end{equation}
The Ricci scalar function is depicted in Fig. \ref{FigRicci} for different values ​​of $\theta$ parameter. As can be seen from the diagram, the Ricci scalar has become infinite at two points $ m+\sqrt{m^2-a^2+n^2} $ and $ m-\sqrt{m^2-a^2+n^2} $ and reaches zero at infinity.

\begin{figure}[!htb]
\centering
\subfloat[]{
\includegraphics[width=0.32\textwidth]{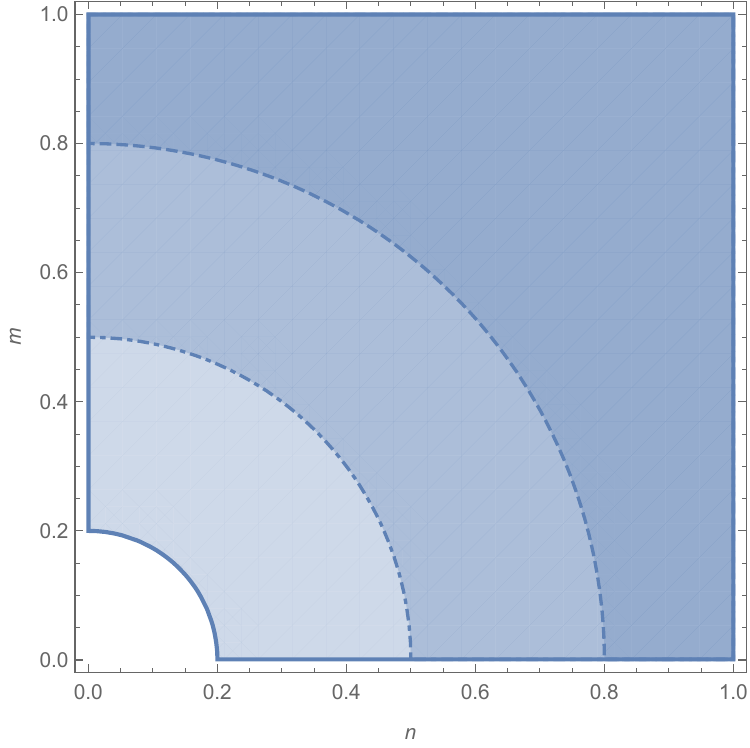}}
\subfloat[]{
\includegraphics[width=0.32\textwidth]{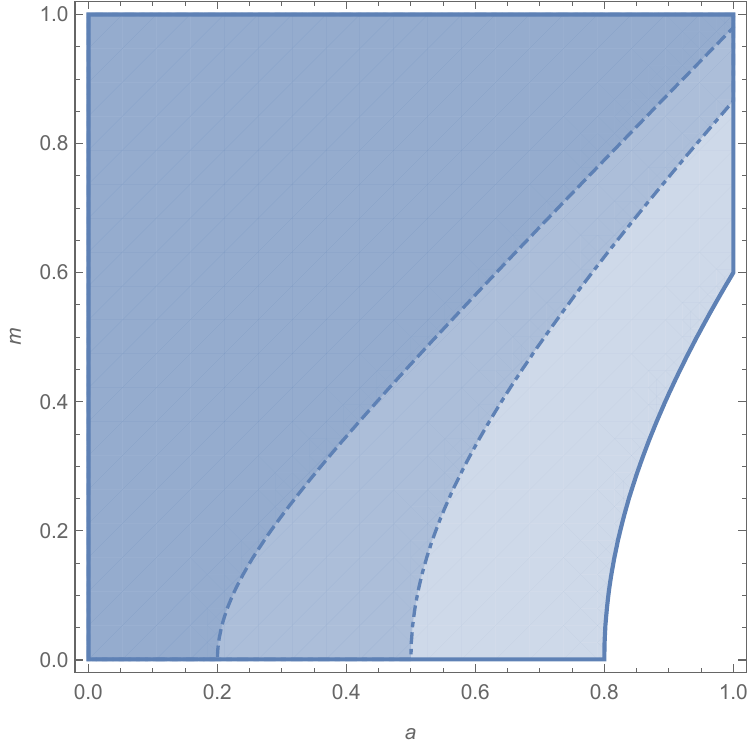}}
\subfloat[]{
\includegraphics[width=0.32\textwidth]{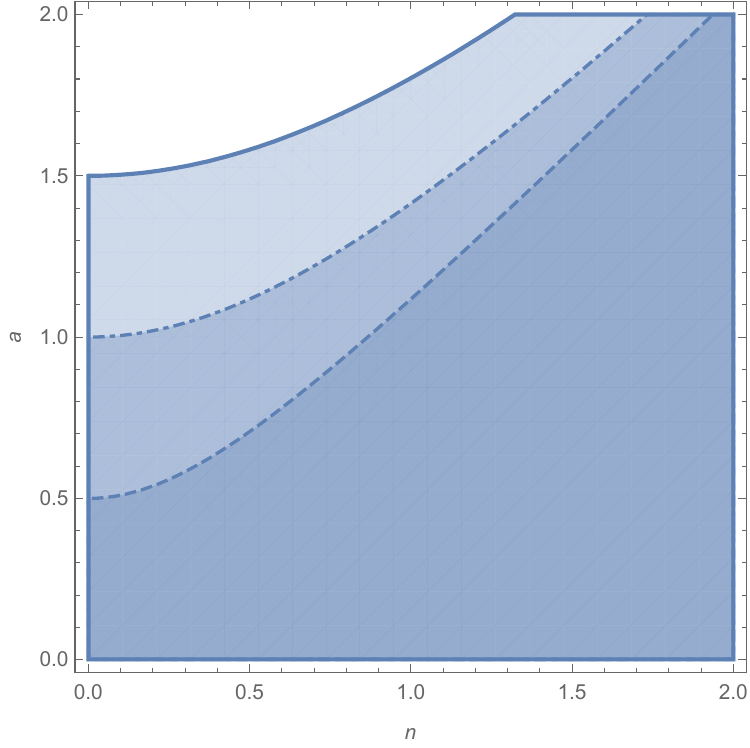}}        
\newline
\caption{The admissible region (denoted by shaded areas) is displayed in (a): $ m -n $ plane for $ \nu=-2$ and $a=0.2$ (continuous Curve), $a=0.5$ (dashdotted Curve) and $a=0.8$ (dashed Curve); (b): $m - a$ plane for $ \nu=-2$ and $n=0.8$ (continuous Curve), $n=0.5$ (dashdotted Curve) and $n=0.2$ (dashed Curve);(c): $a - n$ plane for $ \nu=-2$ and $m=1.5$ (continuous Curve), $m=1.0$ (dashdotted Curve) and $m=0.5$ (dashed Curve).}
\label{Fig1}
\end{figure}

Before studying the optical features of this solution, we would like to explore the allowed regions of parameters in
which a physical solution exists. The admissible parameter space is displayed in Fig. \ref{Fig1}. The white area is where no singularity solution exists. The continuous, dashed, and dash-dotted curves are the boundary for the existence of singularities in the bulk, with extremal singularities sitting on the curve. Looking at Fig. \ref{Fig1}(a), one can find that the admissible region decreases with an increase in the rotation parameter while increasing the NUT charge increases the allowed region (see Fig. \ref{Fig1}(b)). 

\section{RAY-TRACING AND SHADOW BOUNDARY}
\label{Sec2}
\begin{figure}[!htb]
\centering
\includegraphics[width=0.36\textwidth]{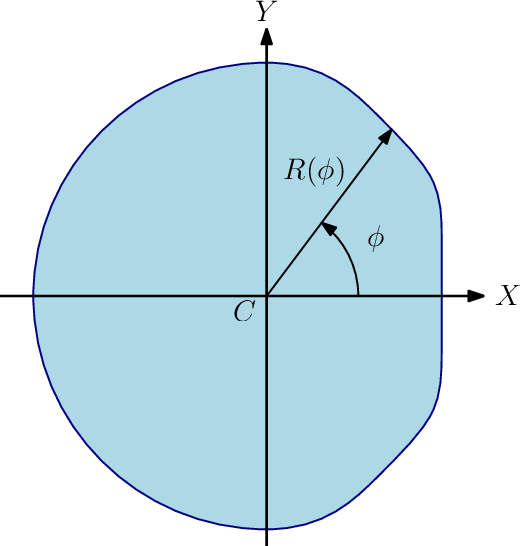}        
\newline
\caption{The function  $ R(\varphi) $ is characterized as the measurement of distance from every point along the boundary of the shadow to the center of that shadow.}
\label{Fig0}
\end{figure}

In this section, we turn our attention to the shadow image of a class of KTNS metrics in \eqref{metric19}. The experimental results reported by EHT provided the first image of a supermassive singularity in the galaxy M87. The image shows a dark part surrounded by a bright ring, which are called the singularity shadow and photon sphere, respectively. The photon sphere surrounding a singularity in spacetime acts as a boundary that distinguishes between geodesics that are drawn into the event horizon and those that manage to escape towards spatial infinity. To study the singularity shadow, we need to solve the null-geodesic equations analytically. In the context of Kerr spacetime, the photon sphere is characterized by an analytical solution; however, for more general spacetimes, the equations governing null geodesics do not yield analytical solutions. Consequently, it becomes necessary to employ numerical methods to accurately compute the shadow. For our case, since the geodesic equations obtained from the metric are not separable, we adopt the numerical ray-tracing method to investigate the shadow of these singularities. We use simulation that has been done in 
\cite{70--ghasemi2015shadows, 71--ghasemi2016note} 
and  address the equations utilizing a variant of the Runge-Kutta-Nystrom method, which incorporates an adaptive step size and mechanisms for error control. The initial conditions for the photon must first be established in the image plane of the distant observer, who is positioned at a distance $D$ from the singularity and at an inclination angle $i$. The image plane as perceived by the distant observer is represented using a Cartesian coordinate system denoted as $ (X,Y,Z) $, whereas the metric is characterized by the coordinates $ (x,y,z) $. In this context, we consider the image plane to be situated at the position defined by the equation $Z=0$, with the Z-axis oriented perpendicularly to the image plane. The relationship between the two Cartesian coordinates is established as follows
\begin{eqnarray}
x &=& D \sin i - Y \cos i + Z \sin i \, , \nonumber\\
y &=& X \, , \nonumber\\
z &=& D \cos i + Y \sin i + Z \cos i \, .
\end{eqnarray}
The transformation from Cartesian metric coordinates to spherical coordinates can be achieved through the standard conversion process
\begin{eqnarray}
r &=& \sqrt{x^2 + y^2 + z^2} \, , \nonumber\\ 
\theta &=& \arccos \left(\frac{z}{r}\right) \, , \nonumber\\
\phi &=& \arctan \left(\frac{y}{x}\right) \, .
\end{eqnarray}
Taking into account a photon located at the coordinates $(X_0, Y_0, 0)$, which possesses a uniform initial momentum represented by ${\bf k}_0 = - k_0 \hat{Z}$ and is oriented perpendicular to the image plane, the initial conditions for the photon can be articulated as follows
\begin{eqnarray}\label{eq-m-initial-x}
t_0 &=& 0 \, , \nonumber\\
r_0 &=& \sqrt{X_0^2 + Y_0^2 + D^2} \, , \nonumber\\
\theta_0 &=& \arccos \frac{Y_0 \sin i + D \cos i}{r_0} \, , \nonumber\\
\phi_0 &=& \arctan \frac{X_0}{D \sin i - Y_0 \cos i} \, .
\end{eqnarray}
and the initial equations for the 4-momentum of the photon are as below
\begin{eqnarray}\label{eq-m-initial-k}
k^r_0 &=& - \frac{D}{r_0} |{\bf k}_0| \, , \nonumber\\
k^\theta_0 &=& \frac{\cos i - \left(Y_0 \sin i + D \cos i\right) 
\frac{D}{r_0^2}}{\sqrt{X_0^2 + (D \sin i - Y_0 \cos i)^2}} |{\bf k}_0| \, , \nonumber\\
k^\phi_0 &=& \frac{X_0 \sin i}{X_0^2 + (D \sin i - Y_0 \cos i)^2} |{\bf k}_0| \, , \nonumber\\
k^t_0 &=& \sqrt{\left(k^r_0\right)^2 + r^2_0  \left(k^\theta_0\right)^2 + r_0^2 \sin^2\theta_0  (k^\phi_0)^2} \,.
\end{eqnarray}
It is worth noting that $k^t_0$ is obtained from the condition $g_{\mu\nu}k^\mu k^\nu = 0$. To parametrize the boundary of the shadow, we define the shadow center as follows 
\begin{eqnarray}
X_{\rm center} &=& \frac{\int \int \rho(X,Y) X dX dY}{\int \int \rho(X,Y) dX dY} \,  \nonumber\\
Y_{\rm center} &=& \frac{\int \int \rho(X,Y) Y dX dY}{\int \int \rho(X,Y) dX dY} \, ,
\end{eqnarray}
In this context, we have defined the relationship $ \rho(X,Y)=1 $ to represent points within the shadow and $ \rho(X,Y)=0 $ for those outside of it. Given the symmetry of the shadow about the X-axis, the shorter segment along the X-axis serves as the starting point, denoted as $ \varphi =0 $. The characteristics of the singularity shadow are described by the function $ \frac{R(\varphi)}{R(0)} $, where $ R(\varphi) $ indicates the distance from point C to the boundary at the angle $ \varphi $, and $ R(0) $ represents the value of R when $ \varphi =0 $ (refer to Fig. \ref{Fig0} for further details).

\subsection{Testing corresponding singularities using the M87* observation}
\begin{figure}[!htb]
\centering
\subfloat[ $ \nu=0 $]{
\includegraphics[width=0.31\textwidth]{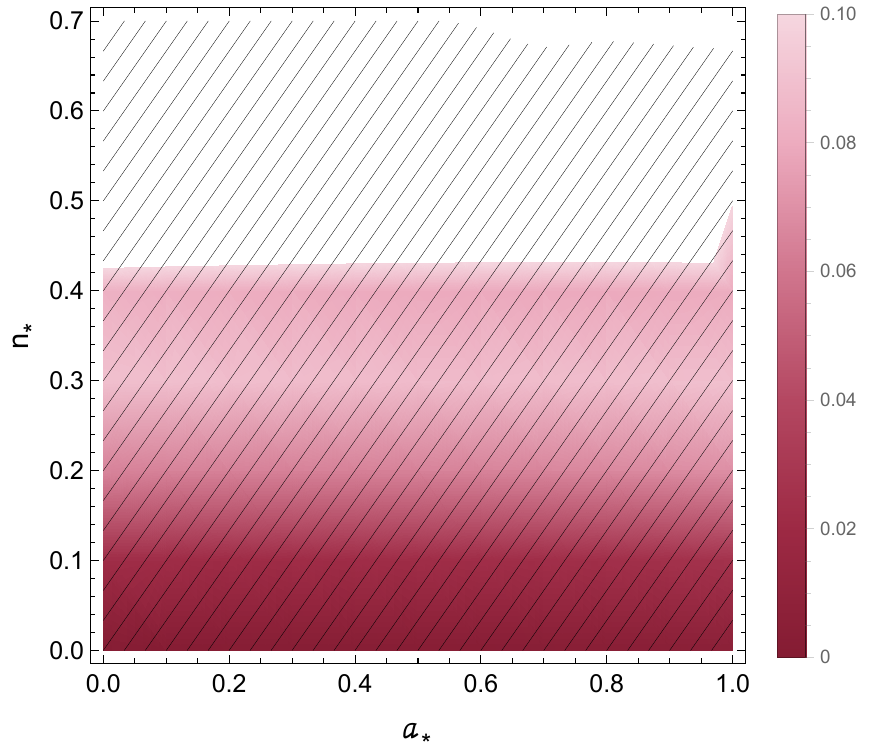}}
\subfloat[$ \nu=-1 $]{
\includegraphics[width=0.31\textwidth]{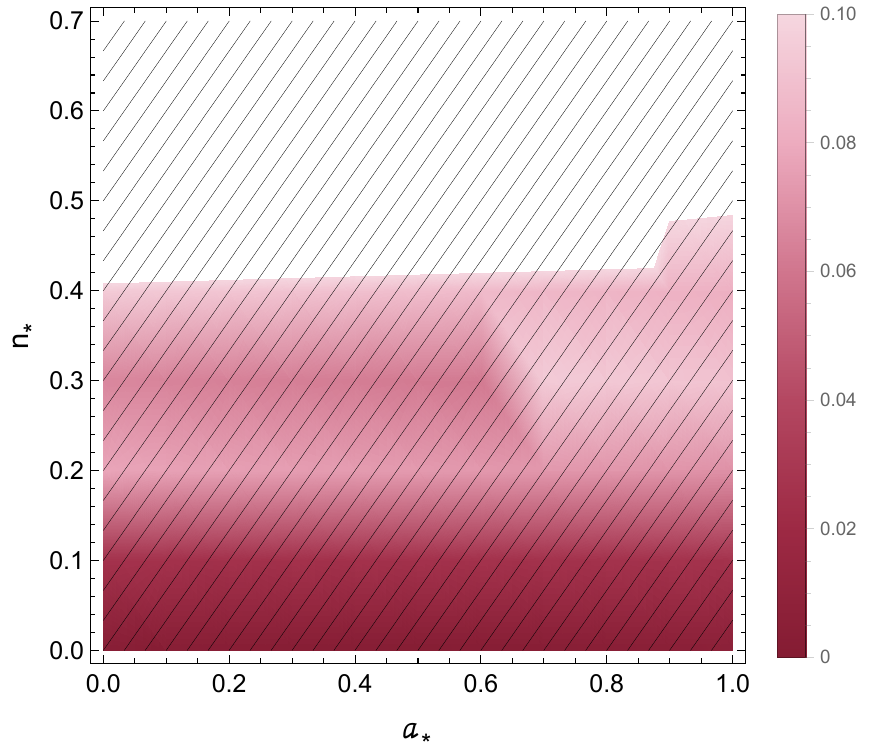}}
\subfloat[$ \nu=-2 $]{
\includegraphics[width=0.31\textwidth]{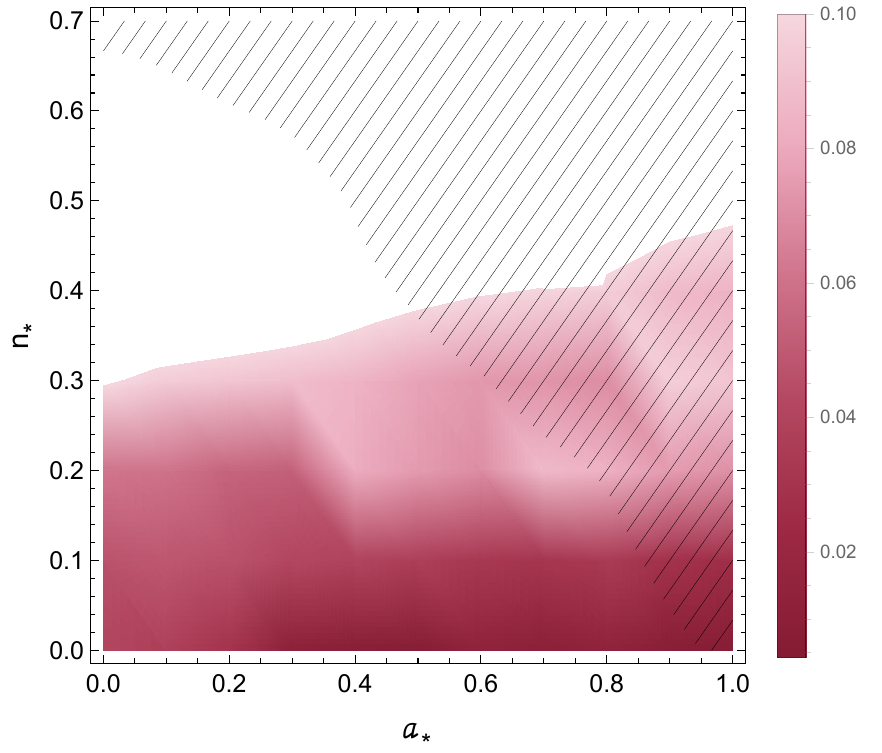}}        
\newline
\subfloat[$ n=0 $]{
\includegraphics[width=0.31\textwidth]{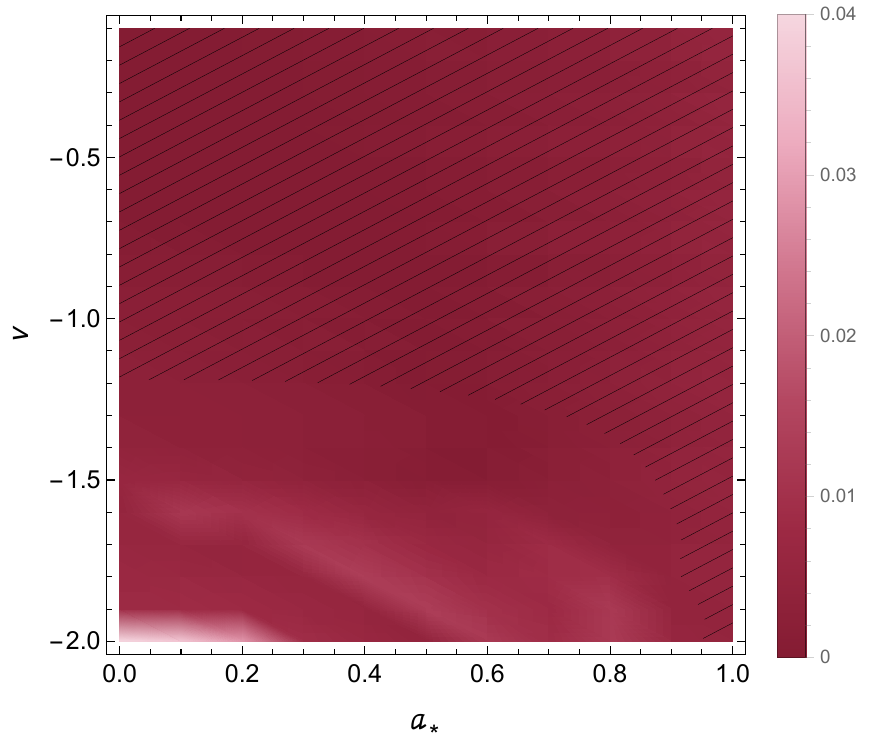}}
\subfloat[$ n=0.1 $]{
\includegraphics[width=0.31\textwidth]{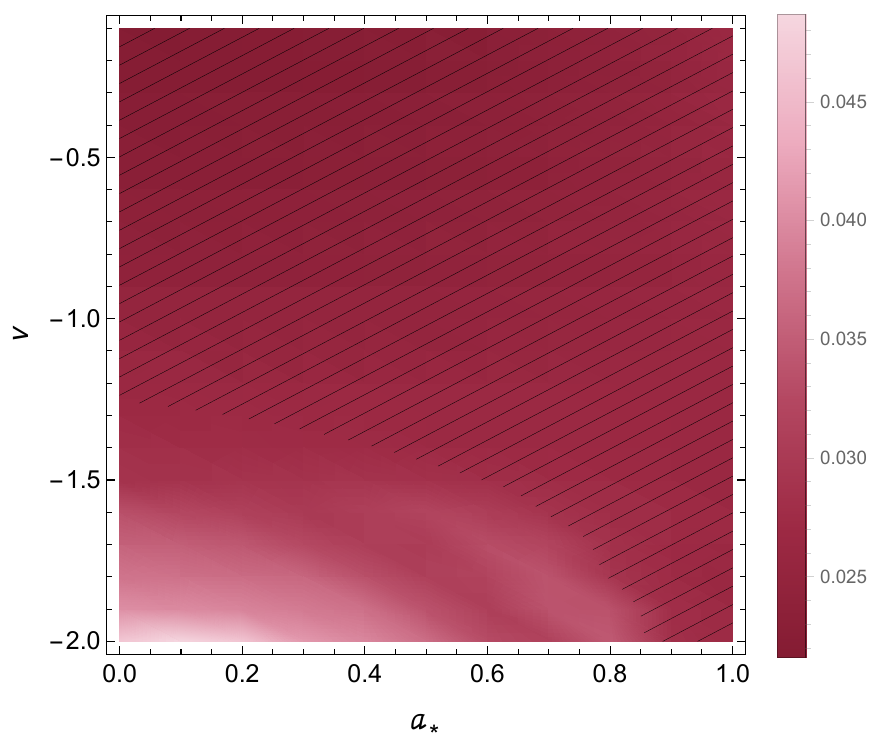}}
\subfloat[$ n=0.3 $]{
\includegraphics[width=0.31\textwidth]{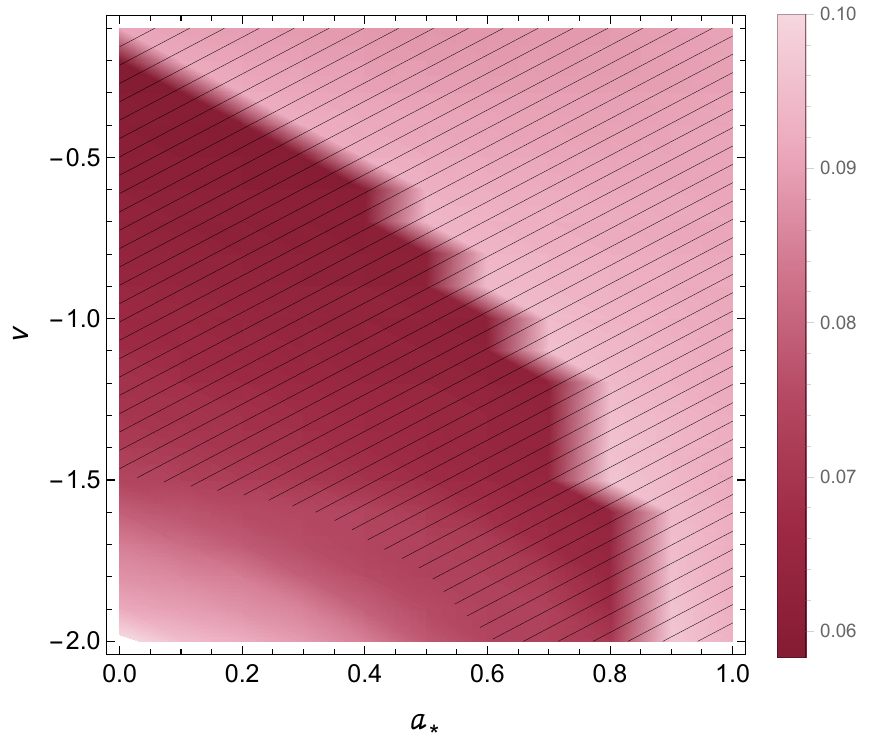}}        
\newline
\caption{The density plot of the circularity deviation $ \Delta C $ with $ \theta=17^{\circ} $  and different values of $ \nu $ and $n$.}
\label{Fig2}
\end{figure}

In this subsection, we examine M87* within the framework of Kerr-Taub-NUT metrics in the presence of a scalar field, and we assess the limitations placed on the metric parameters through the application of EHT observations. The findings from the EHT collaboration have established constraints regarding the angular diameter, the mass of the singularity, and the distance of M87* from Earth, as below
\cite{72--akiyama2019first}
\begin{eqnarray}
\theta_{M87*} &=& (42 \pm 3) \mu as,\\
M_{M87*} &=& (6.5 \pm 0.9) \times 10^{9} M_{\odot},\\
\mathbb{D}_{M87*} &=& 16.8^{+0.8}_{-0.7} Mpc.
\label{EqdM87a}
\end{eqnarray}
In this context, $M_{\odot}$ represents the mass of the Sun. Based on the provided data, the diameter of the shadow, expressed in terms of mass units, adheres to the following condition 
\cite{73-74--bambi2019testing}
\begin{equation}
d_{M87^{*}}\equiv \frac{\mathbb{D}\theta}{M}\approx 11.0 \pm 1.5.
\label{EqdM87b}
\end{equation}
This shows that a KTNS metric can be a candidate for modelling the supermassive hole M87* if the computed shadow satisfies the condition in Eq. \eqref{EqdM87b}. Therefore, this condition will impose a constraint on the parameters of the model. The deviation from circularity is another significant observable linked to the shadow of M87*, which proves valuable in identifying the permissible regions of parameters. Unlike a non-rotating spherically symmetric metric whose shadow boundary is a perfect circle, the shadow structure for rotating solutions has an appearance resembling the dented circular shape, and that is why calculating the deviation from the circular shape of the shadow is nessecay to dtermine the allowed regions of parameters. To determine the circularity of the shadow boundary, it is essential to establish the average radius as below
\cite{73-74--bambi2019testing, 75--banerjee2020silhouette}
\begin{equation}
\bar{\mathbb{R}}^2=\frac{1}{2\pi}\int^{2\pi}_{0} \mathbb{R}(\phi)^2d\phi.
\end{equation}
The expression for $\mathbb{R}(\phi)$ is given by the formula $\mathbb{R}(\phi)=\sqrt{(\alpha-D_c)^2+\beta(\alpha)^{2}}$, where it is specified that $D_c=0$. Additionally, the angle $\phi$ is defined as $\phi=\tan^{-1}\left(\frac{\beta(\alpha)}{\alpha}\right)$. The deviation of the shadow from circularity can be defined as 
\cite{73-74--bambi2019testing, 75--banerjee2020silhouette}
\begin{equation}\label{eq-DeltaC}
{\Delta C=\frac{1}{\bar{\mathbb{R}}}\sqrt{\frac{1}{2\pi}\int^{2\pi}_{0}(\mathbb{R}(\phi)-\bar{\mathbb{R}})^2 d\phi}},
\end{equation}
According to observations made by the EHT collaboration, the deviation $ \Delta C $ from circularity is less than 10\%. To find the allowed region of parameters which are in agreement with EHT data of M87*,  we utilize the observational parameters $ \Delta C $ and the shadow size as described in Eq. \ref{EqdM87b} to impose constraints on the parameters. To achieve this, we have illustrated Fig. \ref{Fig2}, where the colored area indicates the extent of deviation from circularity, while the hatched area delineates the permissible size of the shadow as defined by Eq. \ref{EqdM87b}. Up panels of Fig. \ref{Fig2} depict the deviation $ \Delta C$ in (a-n) plane for different values of $ \nu $. As illustrated in Fig. \ref{Fig2}(a), in the case of the rotating FJNW metric where $\nu=0$, the condition $ \Delta C <0.1$ holds true for values of $n$ less than 0.45. Furthermore, it is observed that the permissible range of the NUT charge, consistent with the EHT data, diminishes as the absolute value of $\nu$ increases. Looking at up panels of this figure, it becomes evident that the sizes of shadow of fast-rotating singularities are more consistent with the EHT data compared to slowly rotating KTNS metrics for large values of $\vert \nu \vert$. Down panels of Fig. \ref{Fig2} illustrate deviation $ \Delta C$ in $(a-\nu)$ plane for different values of $ n $. As see from this figure, the results consistent with EHT data increases as the NUT charge increases.

\subsection{Testing the KTNS metrics using the SgrA* observation}
\begin{figure}[!htb]
\centering
\subfloat[ $ \nu=0 $]{
\includegraphics[width=0.31\textwidth]{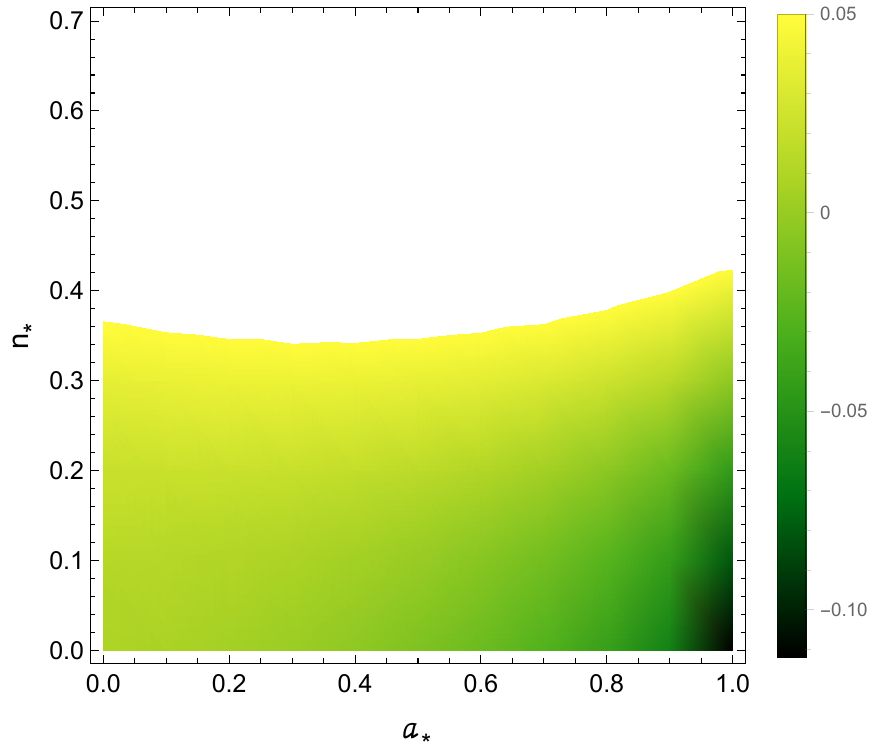}}
\subfloat[$ \nu=-1 $]{
\includegraphics[width=0.31\textwidth]{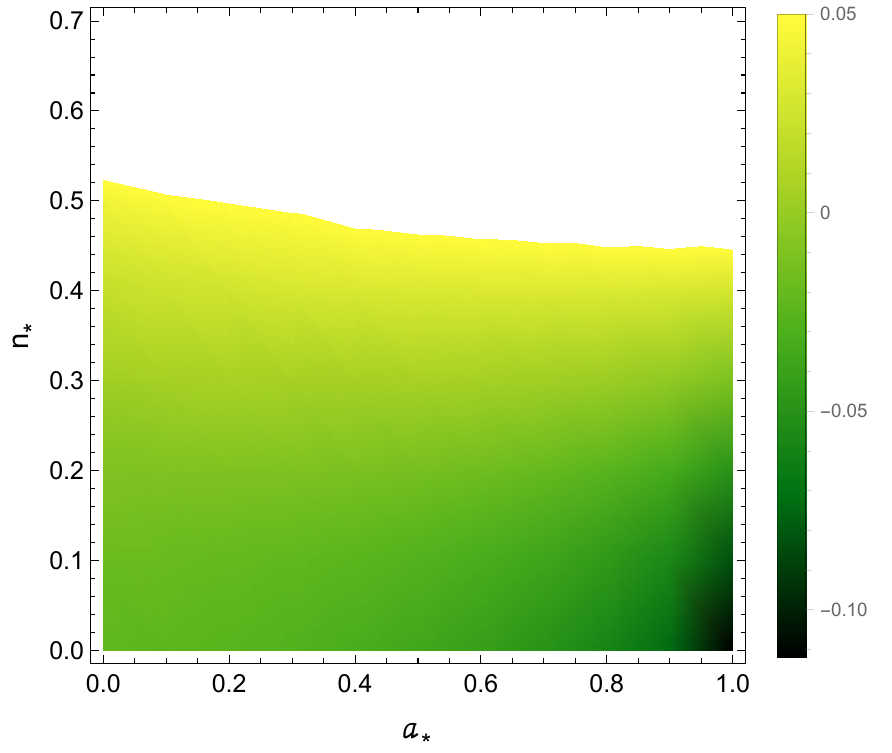}}
\subfloat[$ \nu=-2 $]{
\includegraphics[width=0.31\textwidth]{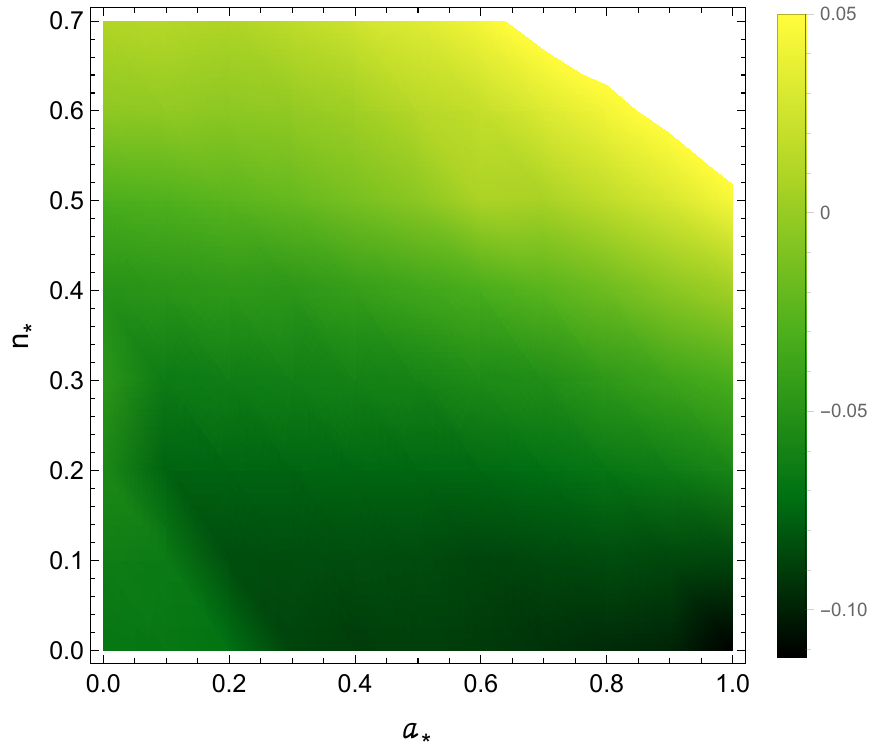}}        
\newline
\subfloat[$ n=0 $]{
\includegraphics[width=0.31\textwidth]{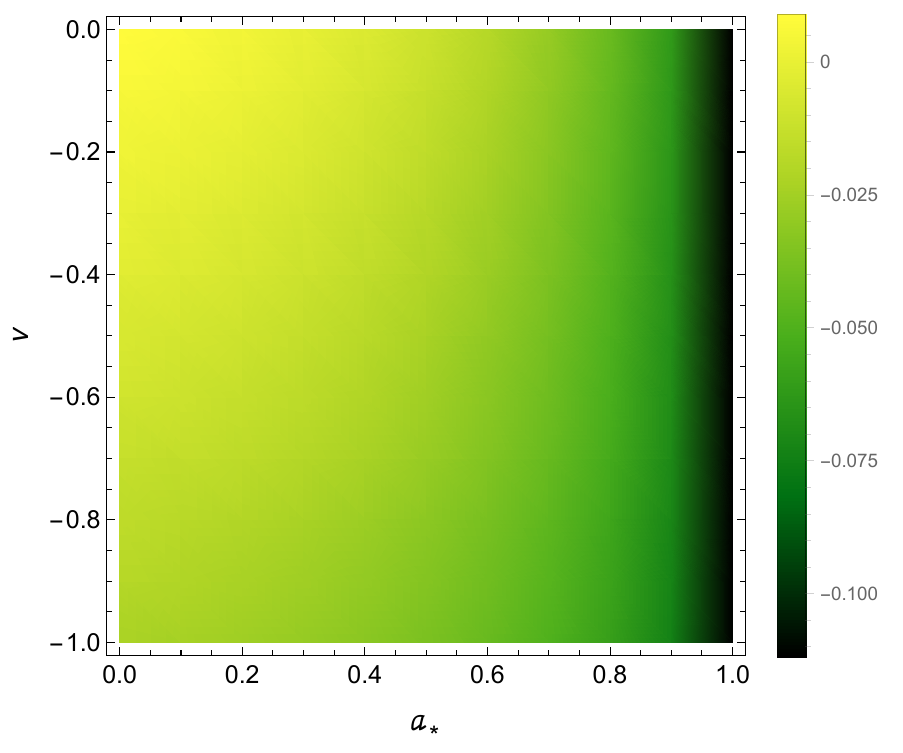}}
\subfloat[ $ n=0.1 $]{
\includegraphics[width=0.31\textwidth]{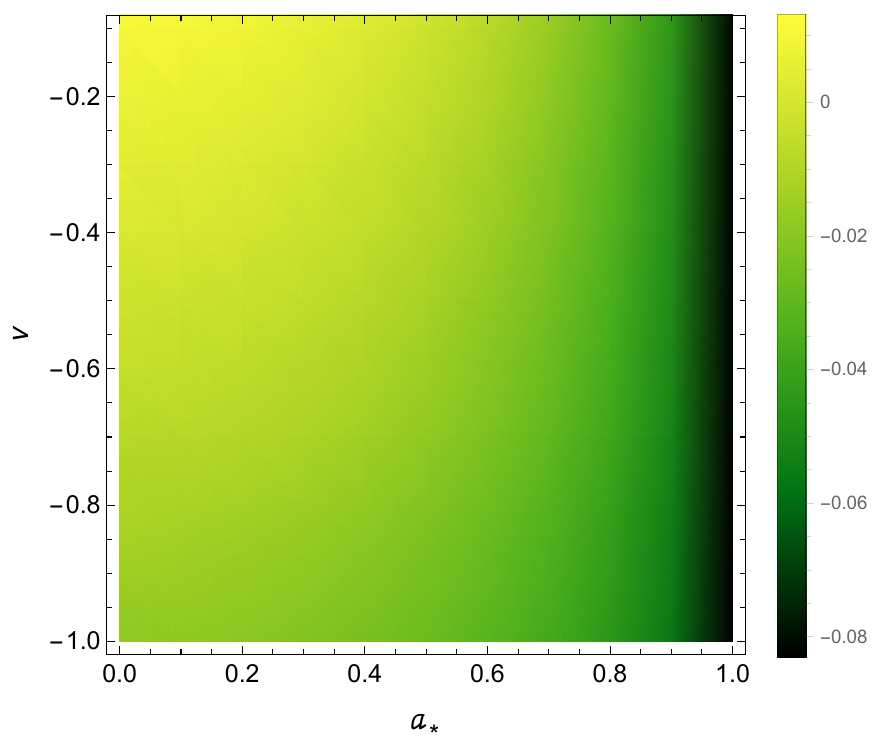}}
\subfloat[$ n=0.3 $]{
\includegraphics[width=0.31\textwidth]{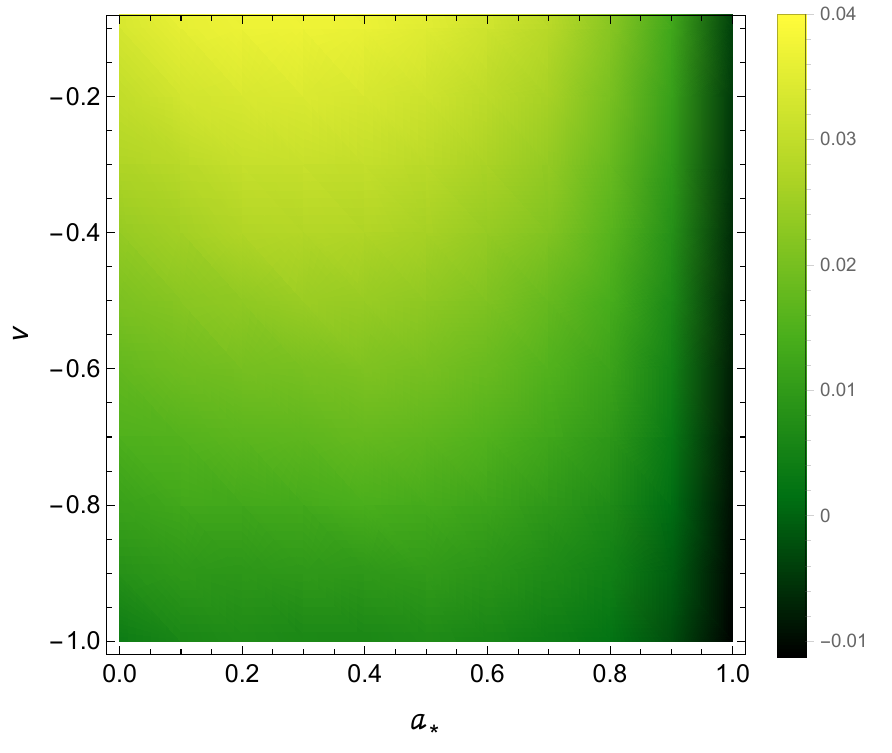}}       
\newline
\caption{The limitations imposed by the results of the EHT regarding the deviation of the Schwarzschild shadow $ \delta = -0.04^{+0.09}_{-0.10}$ (Keck) based on SgrA* observations at inclinations $50^{\circ}$ for different values of $ \nu $ and $ n $.}
\label{Fig4}
\end{figure}

\begin{figure}[!htb]
\centering
\subfloat[$ \nu=0 $]{
\includegraphics[width=0.31\textwidth]{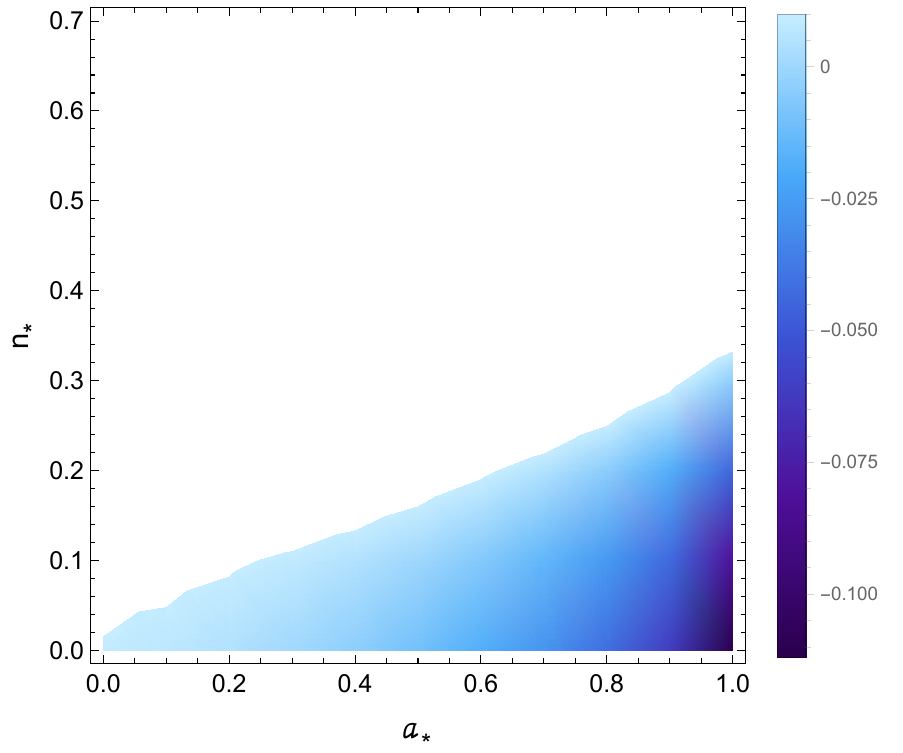}}
\subfloat[$ \nu=-1 $]{
\includegraphics[width=0.31\textwidth]{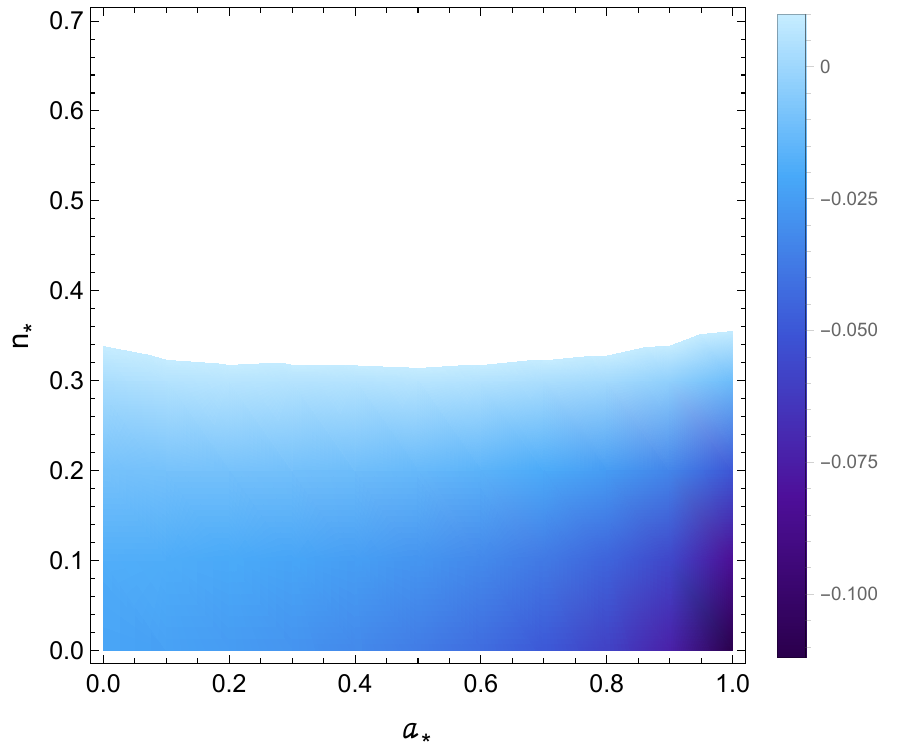}}
\subfloat[$ \nu=-2 $]{
\includegraphics[width=0.31\textwidth]{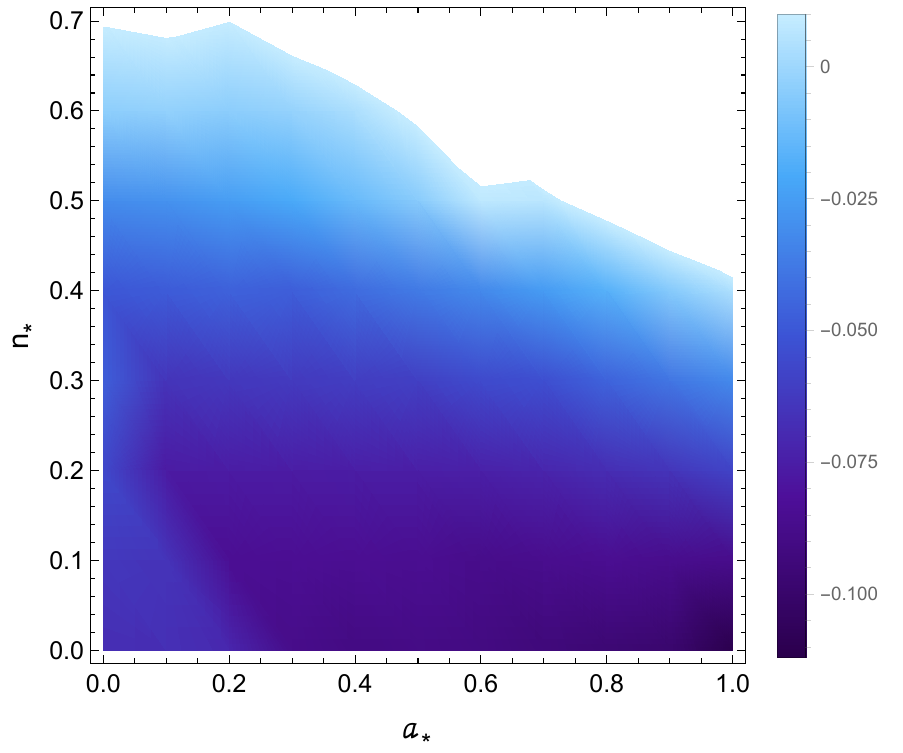}}        
\newline
\subfloat[$ n=0 $]{
\includegraphics[width=0.31\textwidth]{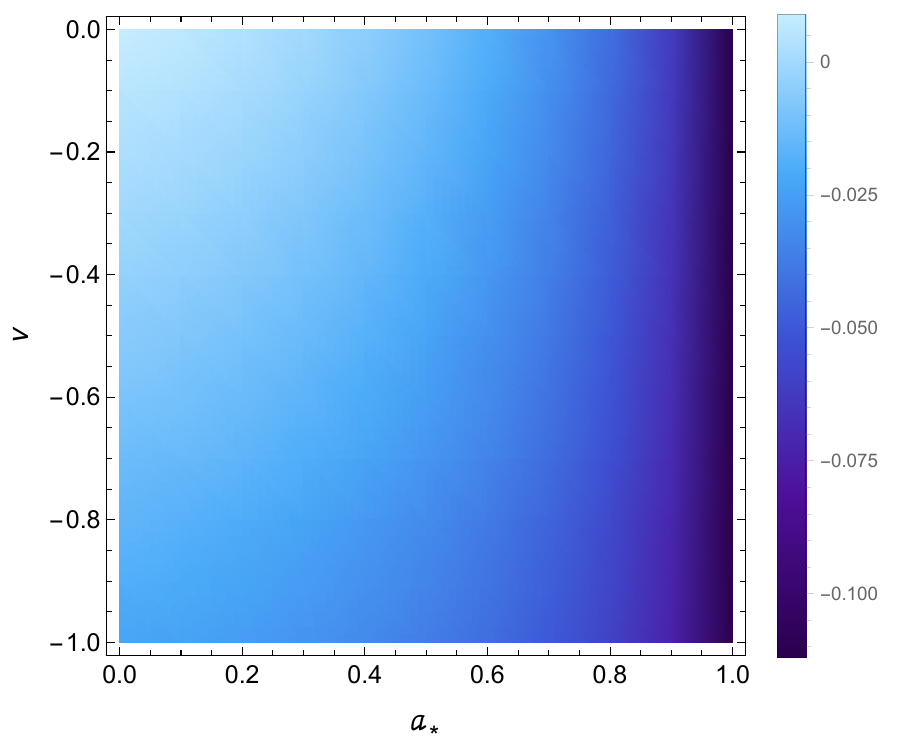}}
\subfloat[$ n=0.1 $]{
\includegraphics[width=0.31\textwidth]{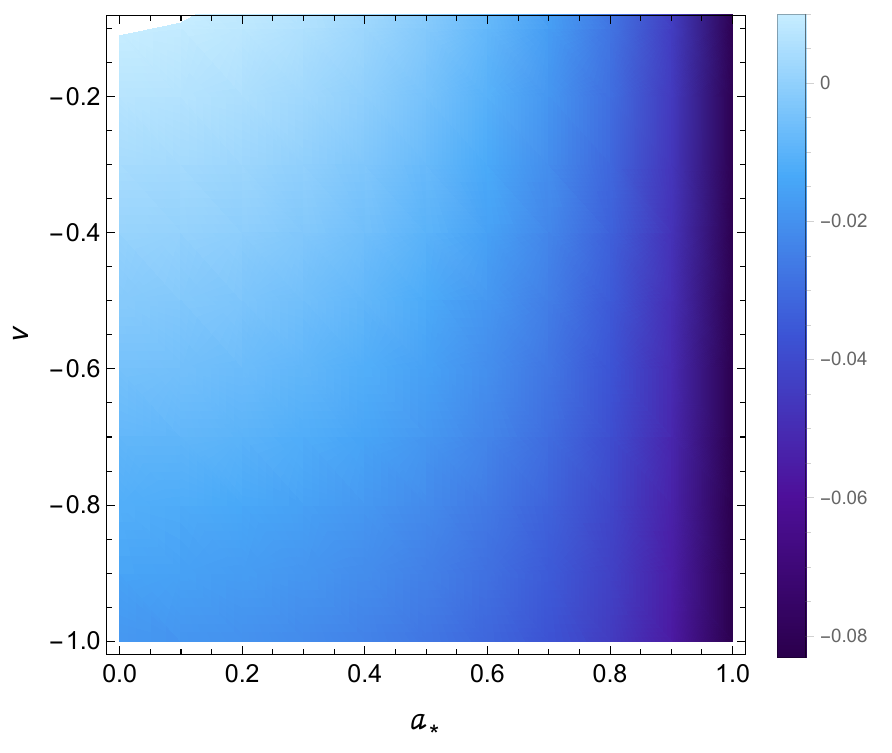}}
\subfloat[$ n=0.3 $]{
\includegraphics[width=0.31\textwidth]{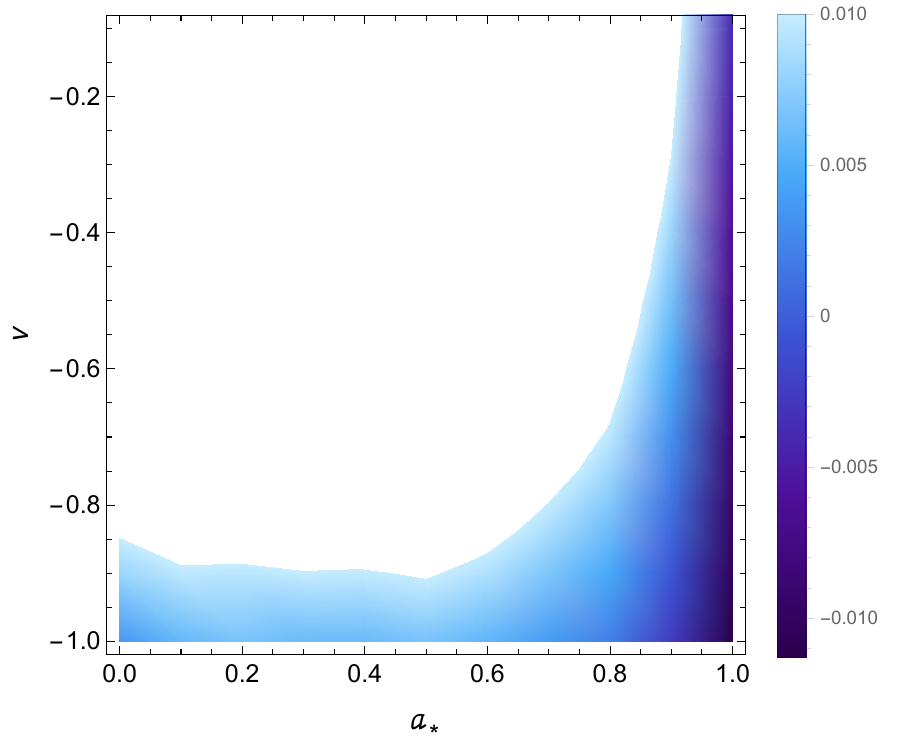}}         
\newline
\caption{The limitations imposed by the results of the EHT regarding the deviation of the Schwarzschild shadow $ \delta = -0.08^{+0.09}_{-0.09}$ (VLTI) based on SgrA* observations at inclinations $50^{\circ}$  for different values of $ \nu $ and $ n $.}
\label{Fig6}
\end{figure}

We are currently limiting the metric parameters by utilizing observations of SgrA* from the EHT. In order to investigate the characteristics of astrophysical supermassive objects, utilizing shadow images of the Sgr A* proves to be more effective than relying on the image captured by M87*. This is primarily because the mass of Sgr A* occupies an intermediary position between the stellar massive objects detected by LIGO and the M87*, thereby enabling a probing of a notably distinct curvature scale. When evaluating the Sgr A*, the mass is typically assessed to be $ M=(4.3 \pm 0.013) \times 10^{6} M_{\odot} $, whereas the distance from Earth is generally approximated as $\mathbb{D}= 8277\pm 33 pc $
\cite{76--akiyama2022first}. 
The EHT focused on Sgr A* has successfully measured the angular diameter of the emission ring, determined to be $\theta_{d} = (51.8 \pm 2.3)\mu as$. Additionally, it has offered an estimation of the shadow diameter, which is calculated to be $\theta_{sh} = (48.7 \pm 7)\mu as$
\cite{77--akiyama2022first}. 
Using these reported numbers, the shadow diameter for Sgr. A* is obtained as
\begin{equation}
d_{Sgr. A*}\equiv \frac{\mathbb{D}\theta_{sh}}{M}\approx 9.5 \pm 1.4.
\label{EqdSgrb}
\end{equation}
The most recent findings from the EHT concerning the observation of Sgr A* have employed the fractional deviation parameter $ \delta $ to assess the disparity between the predicted shadow diameter and that of a Schwarzschild black hole in the following manner
\cite{76--akiyama2022first, 77--akiyama2022first}
\begin{equation}
\delta =\frac{\theta_{sh}}{\theta_{sh,Sch}}-1
\label{Eqdelta}
\end{equation}
EHT utilized two separate previous measurements of the shadow size obtained from the Very Large Telescope Interferometer (VLTI) and Keck observations, thereby determining the subsequent constraints on the fractional deviation parameter $ \delta $
\cite{76--akiyama2022first, 77--akiyama2022first}
\begin{equation}
\delta=\left\{
\begin{array}{ll}
-0.08^{+0.09}_{-0.09}  & \;\;(\mbox{VLTI}) \\
-0.04^{+0.09}_{-0.10}  & \;\;(\mbox{Keck})
\end{array}
\right..
\end{equation}
The fractional deviation parameter is therefore constrained within the intervals of $-0.14 \leq\delta \leq 0.05$ for Keck and $-0.17 \leq\delta \leq 0.01$ for VLTI. Fig. \ref{Fig4} illustrates the fractional deviation parameter $ \delta$ across various values of $ \nu $ and $ n $ with Keck constraint. As illustrated in Fig. \ref{Fig4}(a), it is evident that for the rotating FJNW metric, where $ \nu=0 $, $ \delta$ remains within the confines of the Keck bound for $ n\leq 0.41 $, whereas this allowed region increases as $ \vert \nu \vert $ increases.  According to down panels of Fig. \ref{Fig4},  $  \delta <0$ for very large values of the rotation parameter, meaning that the Schwarzschild black hole shadow is bigger than the shadow of fast-rotating KTNS massive objects. Whereas, for slowly rotating KTNS metric, the shadow is larger than that of Schwarzschild black hole. Looking closely at this figure,  one notices that the value of deviation parameter increases as the NUT charge increases. This reveals the fact that for rotating metrics with large NUT charge, the KTNS shadow is larger than the Schwarzschild black hole shadow.
In Fig. \ref{Fig6}, we have studied the behavior of $  \delta$ with VLTI constraint. As we see from Fig. \ref{Fig6}(a), for rotating FJNW metric, $  \delta$ lies within the VLTI bound for $ n\leq 0.34 $ and the allowed region increases with an increase in $ \vert \nu \vert $. Whereas, the allowed region related to $ \nu $ decreases by increasing NUT charge $ n $ (see down panels of Fig. \ref{Fig6}). 

\section{Weak Deflection Angle}
\label{Sec4}
\begin{figure}[!htb]
\centering
\subfloat[$ n=0.1 $  and $ \nu=-1 $]{
\includegraphics[width=0.32\textwidth]{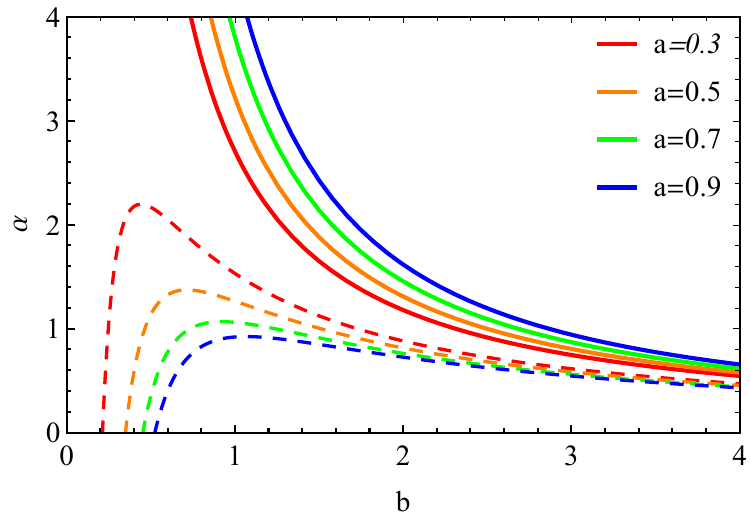}}
\subfloat[$ a=0.8 $ and $ \nu=-1 $]{
\includegraphics[width=0.32\textwidth]{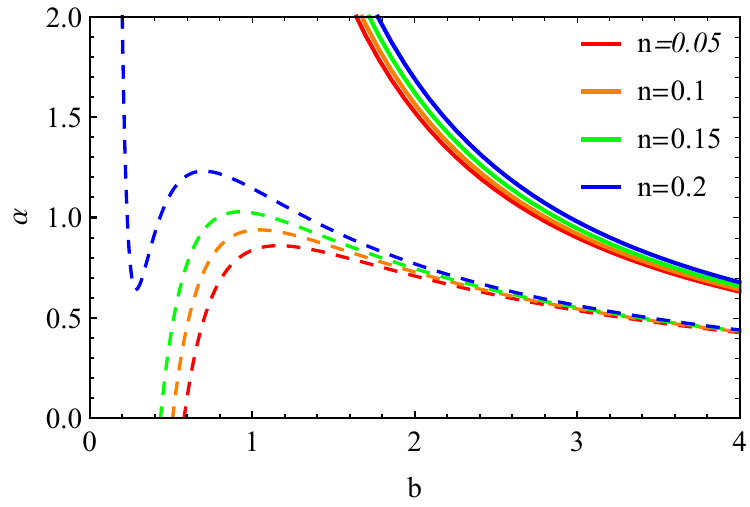}}
\subfloat[$ a=0.5 $ and $ n=0.1 $]{
\includegraphics[width=0.32\textwidth]{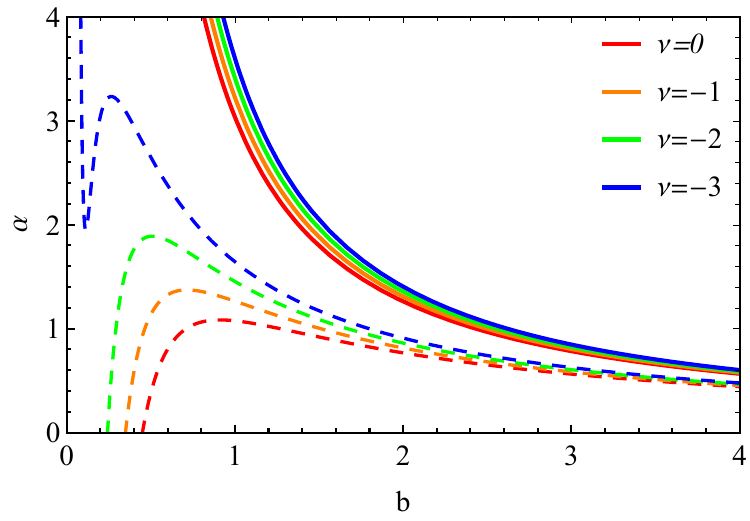}}        
\newline
\caption{Deflection angle $ \alpha_{pro} $ (continuous curves) and $ \alpha_{retro} $ (dashed curves) vs impact parameter $ b $ for various values of $ a $, $ n $ and $ \nu $.}
\label{Fig9}
\end{figure}

In this analysis, we determine the deflection angle of photons traversing the vicinity of mass distribution a KTNS massive object by employing the Gauss-Bonnet theorem. Given that the spacetime described by equation (\ref{metric19}) is axisymmetry, we can utilize the OIA method
\cite{54--ono2017gravitomagnetic}
to ascertain the deflection angle. Typically, the line element for stationary, axisymmetric spacetime in polar coordinates $(r, \theta)$ can be represented as follows
\begin{equation}
ds^{2}=-A(r,\theta)dt^{2}-2H(r,\theta)dt d\phi +B(r,\theta)dr^{2}+C(r,\theta)d\theta^{2}+D(r,\theta)d\phi^{2}
\end{equation}
From Ref. 
\cite{80--asada2000can}, 
we can define the generalized optical metric ($ \gamma_{ij} $) by solving the null condition $ (ds^{2}=0) $ for $  dt$ as 
\begin{equation}
dt=\sqrt{\gamma_{ij}dx^{i}dx^{j}}+\beta_{i}dx^{i}
\end{equation}
where, the indices $i$ and $j$ take on values from 1 to 3, while the functions $ \gamma_{ij} $ and $ \beta_{i} $ are specified as follows
\cite{54--ono2017gravitomagnetic}
\begin{eqnarray}
\gamma_{ij}dx^{i}dx^{j} &\equiv & \frac{B(r,\theta)}{A(r,\theta)}dr^{2}+\frac{C(r,\theta)}{A(r,\theta)}d\theta^{2}+\frac{A(r,\theta)D(r,\theta)+H^{2}(r,\theta)}{A^{2}(r,\theta)}d\phi^{2},\nonumber \\
\beta_{i}&\equiv & -\frac{H(r,\theta)}{A(r,\theta)}d\phi .
\end{eqnarray}
The angle at which light is deflected can be defined as
\begin{equation}
\alpha \equiv \Psi_{R}-\Psi_{S} +\Phi_{RS}
\label{alpha}
\end{equation}
Here, $ \Psi_{R} $ and $ \Psi_{S} $ represent the angles measured at the receiver's location and the source's location, respectively. Furthermore, the expression $ \Phi_{RS} \equiv \Phi_{R}-\Phi_{S}$ defines the difference between the angular coordinates of the receiver, denoted as $ \Phi_{R}$, and the source, represented as $ \Phi_{S}$. The Gauss-Bonnet theorem provides an invariant definition for the deflection angle by identifying appropriate positions for both the receiver and the source, ensuring that the endpoints of the light rays are situated in Euclidean space
\begin{equation}
\alpha=-\iint_{_{R}^{\infty }\square _{S}^{\infty
}}KdS+\int_{S}^{R}\kappa _{g}d\ell .  \label{GB-axial}
\end{equation}
In this context, $ K$ represents the Gaussian curvature within the optical domain, while $ \kappa_{g} $ signifies the geodesic curvature associated with the light ray. The symbols $ dS $ and $ d\ell $ represent the infinitesimal area element of the surface and the element of arc length, respectively. The Gaussian curvature, under the weak field approximation, is computed as follows
\cite{81--ono2018deflection}
\begin{equation}
K=\frac{R_{r\phi r\phi}}{\gamma}=\frac{1}{\sqrt{\gamma}}\Big[\frac{\partial}{\partial\phi}
\Big(\frac{\sqrt{\gamma}}{\gamma_{rr}}\Gamma^{\phi}_{~rr}\Big)
-\frac{\partial}{\partial r}
\Big(\frac{\sqrt{\gamma}}{\gamma_{rr}}\Gamma^{\phi}_{~r\phi}\Big)\Big], 
\label{K}
\end{equation}
where $ \gamma=det(\gamma_{ij}) $.  Since $ \gamma $ is independent of $ \phi $, the first term of Eq. (\ref{K}) does not contribute. To execute the surface integral of the Gaussian curvature as presented in Eq. (\ref{GB-axial}), it is essential to identify the boundaries of the integration domain. To do so, we obtain the light orbit equation by the following relation 
\cite{54--ono2017gravitomagnetic}
\begin{equation}
\left(\frac{d u}{d\phi} \right)^{2}=F(u)
\end{equation}
with
\begin{equation}
F(u)=\frac{u^{4}\left( A(u)D(u)+H(u)^{2}\right)\left(D(u)-2 H(u)b-A(u)b^{2} \right)  }{B(u)\left( H(u)+A(u)b^{2}\right)^{2} }
\end{equation}
where $ u=\frac{1}{r} $ and the impact parameter is denoted as $ b $. Within the framework of the weak-field approximation,
\begin{equation}
u=\frac{\sin \phi }{b}+\frac{m(1+\cos ^{2}\phi )}{b^{2}}-\frac{2am}{b^{3}}.
\end{equation}
The surface integral of the Gaussian curvature, as presented in Eq. (\ref{GB-axial}), is computed as below utilizing the aforementioned solution for the photon orbit 
\begin{eqnarray}
-\iint_{_{R}^{\infty }\square _{S}^{\infty }}KdS &=&\int_{\phi_{S}}^{\phi _{R}}\int_{0}^{u}-\frac{K \sqrt{\gamma}}{u^{2}}dud\phi \\ \nonumber
&=&-\frac{\mathcal{A}}{4b^{2}}\left( \cos ^{-1}bu_{S}+\cos ^{-1}bu_{R}\right) +\frac{2m}{b} \left[\sqrt{1-b^{2}u_{S}^{2}} +\sqrt{1-b^{2}u_{R}^{2}}\right]  \\ \nonumber
&+& \frac{\mathcal{A}}{4b}\left[u_{S}\sqrt{1-b^{2}u_{S}^{2}} +u_{R}\sqrt{1-b^{2}u_{R}^{2}} \right] 
\label{intK}
\end{eqnarray}
where
\begin{equation}
\mathcal{A}= 2a^{2}\nu -2(7+\nu)n^{2}
\end{equation}
The calculation of geodesic curvature can be expressed as follows
\cite{54--ono2017gravitomagnetic}
\begin{equation}
\kappa =-\frac{1}{\sqrt{\gamma \gamma^{\theta \theta}}}\beta_{\phi,r}, 
\label{kappa}
\end{equation}
The path integral of $ \kappa $ is computed as
\begin{eqnarray}
\int_{S}^{R}\kappa d\ell &=& \pm  \frac{\mathcal{B}}{2b^{3}}\left( \sin ^{-1}bu_{R}+\sin ^{-1}bu_{S}\right)\pm\frac{2m a}{b^{2}} \left(\sqrt{1-b^{2}u_{R}^{2}} +\sqrt{1-b^{2}u_{S}^{2}}\right)  \\ \nonumber
&\pm & \frac{\mathcal{B}}{2b^{2}}\left(u_{R}\sqrt{1-b^{2}u_{R}^{2}} +u_{S}\sqrt{1-b^{2}u_{S}^{2}} \right)  , 
\label{kappa2}
\end{eqnarray}
in which $ \mathcal{B}= -4an^{2} $. The positive sign indicates the retrograde motion, while the negative sign corresponds to the prograde motion of the photon orbit. By Combining the solutions of both the Gaussian optical curvature integral and the geodesic curvature integral, it is possible to calculate the overall deflection angle of light in the equatorial plane
\begin{eqnarray}
\alpha &=&-\frac{\mathcal{A}}{4b^{2}}\left( \cos ^{-1}bu_{S}+\cos ^{-1}bu_{R}\right) +\frac{2m}{b} \left(\sqrt{1-b^{2}u_{S}^{2}} +\sqrt{1-b^{2}u_{R}^{2}}\right) \\ \nonumber
&+&\frac{\mathcal{A}}{4b}\left(u_{S}\sqrt{1-b^{2}u_{S}^{2}} +u_{R}\sqrt{1-b^{2}u_{R}^{2}} \right) \pm   \frac{\mathcal{B}}{2b^{3}}\left( \sin ^{-1}bu_{R}+\sin ^{-1}bu_{S}\right)\\ \nonumber
&\pm & \frac{2m a}{b^{2}} \left(\sqrt{1-b^{2}u_{R}^{2}} +\sqrt{1-b^{2}u_{S}^{2}}\right) \pm  \frac{\mathcal{B}}{2b^{2}}\left(u_{R}\sqrt{1-b^{2}u_{R}^{2}} +u_{S}\sqrt{1-b^{2}u_{S}^{2}} \right)  , 
\label{alpha}
\end{eqnarray}
Considering $ u_{S}\rightarrow 0 $ and $ u_{R}\rightarrow 0 $, one can obtain the deflection angle for infinite distance limit. It should be noted that the observer and the source are positioned at a considerable distance from the center of the supermassive black hole. The behavior of the deflection angle is depicted in Fig. \ref{Fig9} which verifies that the angle of deflection becomes greater as the rotation parameter, NUT charge and absolute value of $\nu$ increase. This reveals the fact that photons are more deflected from their straight path around fast rotating spacetime. From Figs. \ref{Fig9}(b) and \ref{Fig9}(c), it can be seen that $ \alpha_{pro} =\alpha_{retro}$ for large values of $n$ and $\vert \nu \vert$, revealing the fact that the orbital angular momentum of the photons will be opposite to the central massive object spin in such a situation.

\section{Conclusions}
\label{Sec5}
Research in the field of imaging and the optical characteristics of massive objects has progressed significantly into an extensive study of shadows and lensing in recent years. The observed images related to M87* and Sgr A* at the heart of Milky Way Galaxy are ideal and real laboratories for exploring and testing fundamental physics and gravity within the context of strong field conditions. We derived a new class of KTNS metrics and constrain the parameters of the model using two shadow measurements from M87* and Sgr A*.  We first employed the EHT observations of M87* and estimated the ranges of KTNS parameters to obtain their upper limit constraint. To this purpose, we used the constraint related to two specific observables: the size of the shadow and the deviation from circularity $ \Delta C $, and found the allowed region of parameters. According to our finding, the condition $ \Delta C <0.1$ is met when $n$ is less than 0.5, and the allowed region of the NUT charge decreases with an increase in $\vert \nu \vert$, whereas the allowed region related to $\nu$ increases as the NUT charge increases. Our results also illustrated that for a given value of $\vert \nu \vert$ and $n$, fast-rotating KTNS metrics are better candidates for suppermassive M87* compared to slowly rotating ones. To place limitation on parameters using SgrA* data, we employed the constraints on the observable fraction deviation, denoted as $\delta$, as reported by measurements from Keck and VLTI. Our analyse showed that for rotating FJNW metric ($ \nu=0 $), the value of $ \delta$ is situated within the confines of the Keck bound for $ n\leq 0.41 $, whereas the allowed range increases in the presence of the parameter $ \nu$. In other words, the allowed range of the NUT charge increases with an increase in $ \vert \nu \vert $.  We discovered that for significantly large values of the rotation parameter, the condition $ \delta <0$ holds true, indicating that the shadow produced by rapidly rotating KTNS metrics is smaller than that of the Schwarzschild black hole. Conversely, in the case of slowly rotating KTNS metrics, the shadow is larger than that of the Schwarzschild black hole. Moreover, we noticed that the value of deviation parameter increases as the NUT charge increases. This reveals the fact that for rotating KTNS metrics with large NUT charge, the shadow of the KTNS is more expansive than that of the Schwarzschild black hole. Regarding the constraints with VLTI observations, our finding illustrated that for rotating FJNW metric, the value of $ \delta $ is situated within the constraints established by the VLTI for $ n\leq 0.34 $ and the allowed range increases as  $ \vert \nu \vert $ increases. Whereas, the allowed range of $ \nu $ decreases by increasing the NUT charge $ n $.

Lastly, we investigated the parameters' effect on the deflection angle of the photons as they traverse near the center of KTNS metrics. Taking into account the observer and the source positioned at finite distances from the KTNS massive object, we derived the analytical formula for the deflection angle within the weak-field approximation. We illustrated that increasing the rotation parameter, NUT charge and absolute value of $\nu$ increase the deflection angle, meaning that photons are more deflected from their straight path around fast rotating KTNS massive objects. Our analysis also showed that the orbital angular momentum of the photons will be opposite to the KTNS spin in the presence of the stronger NUT charge and scalar field.

\begin{acknowledgements}
M. G.-N. acknowledges the support from the CAS Talent Program and the Xinjiang Tianchi Talent Program.
\end{acknowledgements}

\bibliography{references}{}

\end{document}